\documentclass[fleqn,usenatbib]{mnras}

\usepackage{newtxtext,newtxmath}

\usepackage[T1]{fontenc}

\DeclareRobustCommand{\VAN}[3]{#2}
\let\VANthebibliography\thebibliography
\def\thebibliography{\DeclareRobustCommand{\VAN}[3]{##3}\VANthebibliography}

\usepackage{graphicx}	
\usepackage{amsmath}	
\usepackage{subcaption}
\usepackage{tabularx}
\usepackage{float}

\title[GRMHD simulations of accretion flows with multiple loops]{Two-Temperature GRMHD Simulations of Black Hole Accretion Flows with Multiple Magnetic Loops}

\author[]{
Hong-Xuan Jiang,$^{1, 2}$\thanks{E-mail: hongxuan\_jiang@sjtu.edu.cn}
Yosuke Mizuno,$^{1, 3, 4}$\thanks{E-mail: mizuno@sjtu.edu.cn}, Christian M. Fromm$^{5,4,6}$, Antonios Nathanail$^{7,4}$
\\
$^{1}$Tsung-Dao Lee Institute, Shanghai Jiao Tong University, Shengrong Road 520, Shanghai, 201210, People's Republic of China\\
$^{2}$College of Physics, Sichuan University, Chengdu, 610065, People's Republic of China \\
$^{3}$School of Physics and Astronomy, Shanghai Jiao Tong University, 
800 Dongchuan Road, Shanghai, 200240, People’s Republic of China \\
$^{4}$Institut f\"ur Theoretische Physik, Goethe-Universit\"at Frankfurt, Max-von-Laue-Stra{\ss}e 1, D-60438 Frankfurt am Main, Germany \\
$^{5}$Institut f\"ur Theoretische Physik und Astrophysik, Universit\"at W\"urzburg, Emil-Fischer-Str. 31, D-97074 W\"urzburg, Germany \\
$^{6}$Max-Planck-Institut f\"ur Radioastronomie, Auf dem H\"ugel 69, D-53121 Bonn, Germany \\
$^{7}$Department of Physics, National and Kapodistrian University of Athens, Panepistimiopolis, 15783 Zografos, Greece
}

\date{Accepted XXX. Received YYY; in original form ZZZ}

\pubyear{2022}

\begin{document}
\label{firstpage}
\pagerange{\pageref{firstpage}--\pageref{lastpage}}
\maketitle

\begin{abstract}
We have performed a series of two-dimensional two-temperature general relativistic magnetohydrodynamic simulations of magnetized accretion flows initiated from tori with different sizes and poloidal magnetic loop polarities. 
In these two temperature simulations, we trace the process of heating electrons through turbulence and reconnection, most of the time these electrons are trapped in plasmoids.
We found that the accretion process strongly depends on the size of the magnetic loops. The accretion flows never reach the magnetically arrested (MAD) regime in small loop cases. Interaction between magnetic field with different polarities dissipates and decreases the efficiency of magneto-rotational instability. The dependency on the wavelength of the loops places a lower limit on the loop size.
In the large loop cases, after reaching a quasi-steady phase, a transition from Standard And Normal Evolution (SANE) flow to MAD flow is observed. The transition of the accretion state and the transition time depends on the initial loop wavelength. 
The formation of plasmoids strongly depends on the size of the magnetic loops.
The frequent magnetic reconnection between the magnetic loops is responsible for the formation of most of the plasmoids. For some plasmoids, Kelvin-Helmholtz and tearing instabilities are coexisting, showing another channel of plasmoid formation. 
The simulations present that electrons in the plasmoids are well-heated up by turbulent and magnetic reconnection.
Different properties of plasmoid formation in different magnetic field configurations provide new insights for the understanding of flaring activity and electron thermodynamics in Sgr\,A$^*$.
\end{abstract}




\section{Introduction}

Both Sagittarius A$^*$ (Sgr\,A$^*$) and Messier\,87 (M\,87) are categorized as low-luminosity Active Galactic Nuclei (LLAGNs) \citep[e.g.,][]{Yuan2014, 2008ARA&A..46..475H}.
Their accretion rates are well below the Eddington mass accretion rate, $\dot M \ll 10^{-2} \dot M_{\rm Edd}$ \citep[e.g.,][]{Ho2009, Marrone2007}. 
The central supermassive black holes (SMBHs) with different masses in these two LLAGNs have very different outflow or jet structures. 
The relativistic jet in M\,87 propagates to scales of $\rm kpc$ and has been well-collimated. On the other hand, for Sgr\,A$^*$, there is no clear evidence for the existence of a jet. Although there is some discussion for the possibility of the existence of jet and outflow in Sgr\,A$^*$ \citep{Yusef-Zadeh2020}, their velocities are significantly lower than that in M\,87. An important question is why the outflow properties of these two black holes are different. 
Many previous works have indicated that the accretion flow in M\,87 might be magnetically arrested \citep[e.g.,][]{EHTpaperV, Cruz-Osorio2021, Yuan2022}.
The latest Event Horizon Telescope (EHT) observational result of Sgr\,A$^*$ also favors magnetically arrested accretion (MAD) flows compared with the Standard And Normal Evolution (SANE) and other types of accretion flows \citep{Event2022}. However, none of these models passes all the constraints from EHT and non-EHT observations including variability of light curves \citep{Event2022}.

Recently, \cite{Ressler2020} have performed general relativistic magnetohydrodynamic (GRMHD) simulations fed by approximately 30 Wolf-Rayet (WR) stellar winds which are considered to be the situation in Sgr\,A$^*$.
This simulation has reached a MAD accretion flow state during the long-term evolution.
\cite{Murchikova2022} have compared the structure-function of the 230\,GHz light curve obtained from their simulations with three independent observations. The stellar wind-fed model shows the remarkable correspondence of the sub-millimeter time variability. The initial condition of the stellar wind-fed model obtained from large-scale simulations \citep{10.1093/mnras/stz3605}. The parameters in the magnetized WR stellar wind simulations are also consistent with the observation in \cite{Michail2021}.
Several authors have investigated the flare property of MAD flows \citep[][]{2020MNRAS.497.4999D, 2021MNRAS.502.2023P,  2022ApJ...924L..32R, Scepi2022}. 
MAD accretion flows are able to produce highly energetic plasmoids which may correspond to the observed flares. However, in general, MAD flow produces a strong jet via the Blandford-Znajek (BZ) process \citep{1977MNRAS.179..433B} if black hole spin is high enough. Such a highly relativistic jet is appropriate for the case of M87*, while in Sgr\,A$^*$ the observed jet and outflow have not been observed clearly \citep{Yusef-Zadeh2020}.

\cite{Narayan2012} have used two different initial magnetic configurations to produce SANE and MAD accretion flows. For SANE flows, multiple poloidal magnetic loops with alternating polarity have been implemented to avoid magnetic flux accumulation on the horizon to maintain the SANE flows. While for MAD flows, a single large poloidal loop magnetic field configuration was adopted. It is obvious that the initial magnetic field configuration has a significant effect on the produced accretion flow structure. Recalling the idea of the BZ process \citep{1977MNRAS.179..433B}, the black hole spin and the strength of the magnetic field are essential for the formation of jets in black hole accretion flow systems. 
For the case of SANE flows, the magnetic field is weaker, especially for the jet in the funnel region compared with that in MAD flows. The alternating polarity configuration not only avoids the magnetic field accumulation but also leads to the frequent dissipation between the magnetic loops with different polarities which release magnetic energy. A series of two-dimensional (2D) GRMHD simulations with multi-loop configuration has been performed by \cite{Nathanail2020} and 3D simulations initiated with the multi-loop magnetic field have been also performed by \cite{Chashkina2021, Nathanail2021}. These simulations have shown that multiple-loop magnetic field configuration can produce many plasmoids originating from magnetic reconnection, and suggests the formation of striped jets \citep{Chashkina2021}. The previous simulations of multi-loop simulations \citep{Nathanail2020, Nathanail2021, Chashkina2021} show a highly turbulent accretion flow without a strong and stable jet. 


Thanks to the high angular resolution of an interferometric instrument GRAVITY \citep{2017A&A...602A..94G}, a continuous position-changing compact near-infrared (NIR) emission source is found in the observations of Sgr\,A$^*$. Its centroid shows clear orbital motion which is likely corresponding to the high-state (flares) of Sgr A* \citep{GRAVITY2018b}. The emitting region rotates clockwise at a location of a few gravitational radii away from the black hole \citep{GRAVITY2018b}. 
From the light curve of Sgr\,A$^*$\citep[e.g.,][]{TheGRAVITYCollaboration2020}, the variability of bulk emission has a log-normal distribution which contains the quiescent emission supported by sporadic flares. While the flares correspond to the observed power-law extension of the flux distribution \citep{Abuter2020}. \cite{Michail2021} reports the first simultaneous radio observations of Sgr\,A$^*$ across 16 frequencies between $8\,\rm GHz$ and $10\,\rm GHz$ and found a 20 minutes time lag between the $10 \,\rm GHz$ and $8\,\rm GHz$. They characterized the quiescent and flaring components of Sgr\,A$^*$ with an adiabatic expansion model. 
\cite{Nathanail2022} presented images from different simulation models, and proposed that the missing stable jet structure in the imaging at $43 \,\rm GHz$ and $86\,\rm GHz$, could be a smoking gun for constraining different accretion models among MAD, SANE and alternating multi-loop model. Recent EHT-ALMA Sgr\,A$^*$ observations \citep{2022A&A...665L...6W} provided polarization information of the emission of the hot-spot, including polarized light curve and hot-spot modeling. Their interpretation of the hot spot as the magnetically arrested region with vertical magnetic flux shows a good agreement with the observation result. Theoretical modeling suggests that the accretion structure in Sgr\,A$^*$ is likely to be accretion flows with clockwise rotation in MAD state \citep{2020MNRAS.497.4999D,2021MNRAS.502.2023P}. Although \cite{2022ApJ...930L..12E} favors the MAD state as well, the variability of the light curve at 230~GHz does not agree with the observation. Thus, the detailed model of Sgr\,A$^*$ still remains unclear.

\cite{Yuan2003} have developed semi-analytical models that successfully reproduce the global properties of Sgr\,A$^*$ observations (spectral energy distribution and multi-wavelength variability). However, these models have not included the mechanism of the flares and the matter distribution around the black hole. \cite{Yuan2009} proposes an analytical MHD model for the formation of episodic jets, and numerical simulations of this model have been studied by \cite{2020MNRAS.499.1561Z}.
\cite{2022ApJ...933...55C} have performed three-dimensional (3D) GRMHD simulations of magnetized accretion flows inspired by the episodic jet model. From their simulations, plasmoids close enough to the horizon mainly stay in the accretion disk, while the ones further away from the black hole are ejected outwards by magnetic forces.

Multiple mechanisms have been proposed to explain the formation of plasmoids, e.g. plasmoids instability \citep{Ripperda2021},  pressure and pinch instability \citep{McKinney2006,2018ApJ...868..146N,Chatterjee2019}, Parker instability \citep{2022ApJ...933...55C}, and
 coexistence of Kelvin-Helmholtz (KH) and tearing instabilities \citep{2022ApJ...929...62B}. Basically, plasmoid instability is a kind of tearing instability \citep{2017ApJ...850..142C}, thus we use the tearing instability instead of plasmoid instability in this work. 
One of the key ingredients for the formation of plasmoids via tearing instability is the existence of current sheets which will create when the magnetic field reconnects. Meanwhile, other plasma instabilities such as KH and pinch instabilities also play an essential role in the formation of plasmoids via turbulent motion.

In this study, we perform a series of two-dimensional two-temperature GRMHD simulations of magnetized accretion flows initiated with a geometrically thick torus involving multiple poloidal magnetic loops. By changing the magnetic loop polarity and wavelength, we investigate the properties of accretion flows under different magnetic configurations such as different magnetic field loop sizes and polarity. In particular, we focus on the formation mechanism of plasmoids and their electron-heating properties.

This paper is organized as follows. In Section 2, we provide information on our simulation setup including initial condition and magnetic configuration. In Section 3 the dynamics of the accretion flow are discussed. In Section 4, we focus on the plasmoid properties. We present the electron temperature properties and electron heating in plasmoids by the comparison between two-temperature simulations and parameterized heating prescription by the $R-\beta$ model in Section 5. 
Finally, we provide our conclusions in Section 6.

\section{Numerical Setup} \label{Sec: setup}

\begin{table}
\begin{tabular*}{\columnwidth}{@{\extracolsep{\fill}}llll}
\hline
\multicolumn{1}{l}{Case} & a      & Polarity    & $\lambda_{\rm r}$ \\ 
\hline
\tt M0                        & 0.9375 & Same           & 0                    \\
\tt M6a                       & 0.9375 & Alternating   & 6                \\
\tt M6b                       & 0.9375 & Same          & 6                       \\
\tt M6c                       & 0      & Alternating    & 6                 \\
\tt M6d                       & 0      & Same         & 6                        \\
\tt M20a                      & 0.9375 & Alternating    & 20                \\
\tt M20b                      & 0.9375 & Same          & 20                       \\
\tt M20c                      & 0      & Alternating    & 20                \\
\tt M20d                      & 0      & Same          & 20                       \\
\tt M50a                      & 0.9375 & Alternating    & 50                \\
\tt M80a                      & 0.9375 & Alternating    & 80                \\
\tt M80b                      & 0.9375 & Same           & 80                       \\
\tt M80c                      & 0      & Alternating    & 80                \\
\tt M80d                      & 0      & Same           & 80                       \\
\hline                       
\end{tabular*}
\caption{Parameters of each case. The initial magnetic field is set by Eq.~\ref{Eq: MAD_A} (see  Fig.~ \ref{fig: initial condition} for the initial conditions).}
\label{Table: MAD}
\end{table}
\begin{figure}
    \centering
         \includegraphics[width=\linewidth]{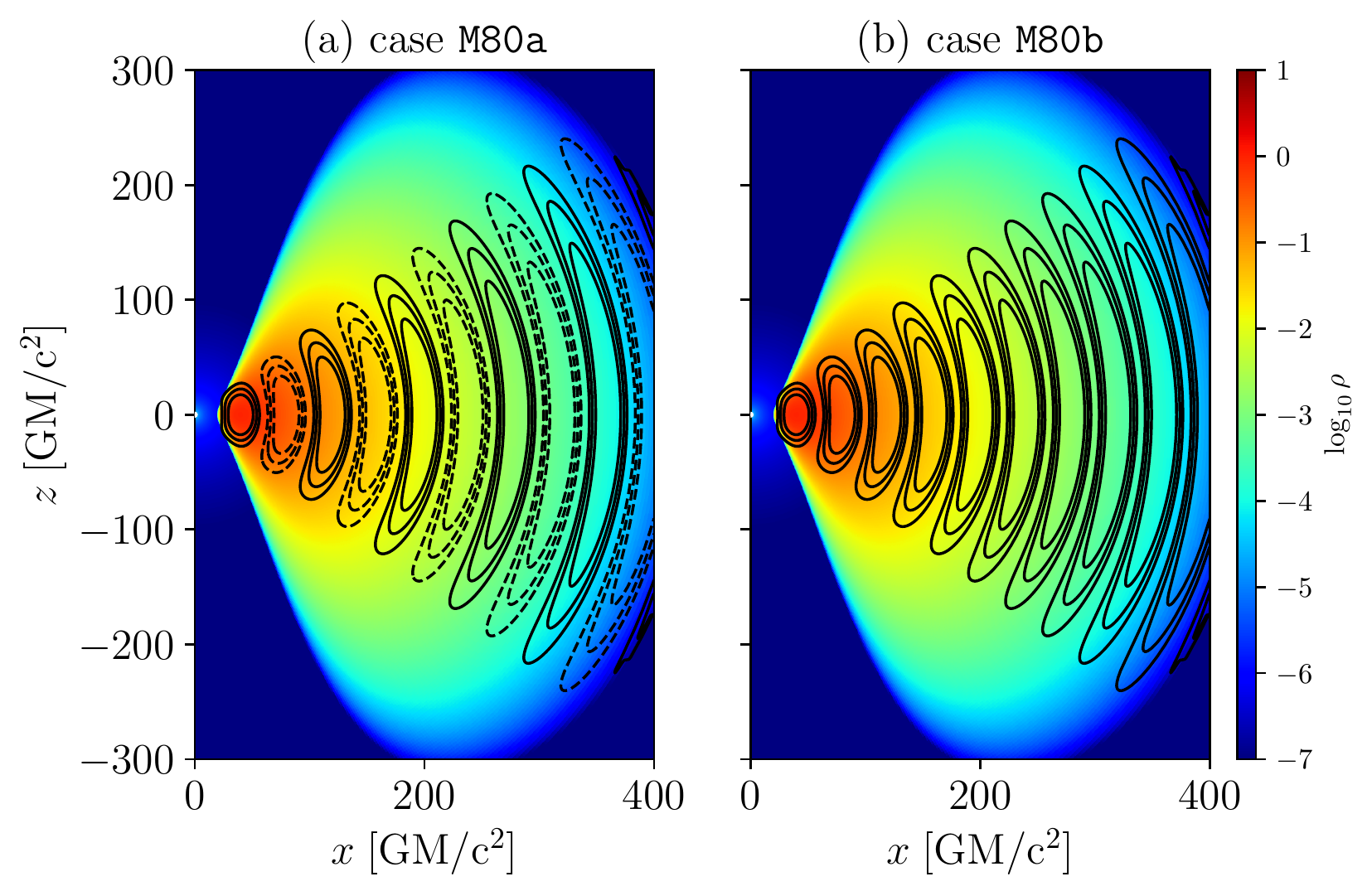}
    \caption{Initial torus profile of logarithmic density (color) and magnetic field configuration (black contours) of two representative cases from the alternating polarity and same polarity groups. Dashed contours represent positive polarity while solid ones are the opposite. The wavelength of the magnetic configuration for the two cases are both 80.}
    \label{fig: initial condition}
\end{figure}
We have performed a series of 2D two-temperature GRMHD simulations of magnetized tori in a black hole spacetime using the $\tt BHAC$ code \citep{Porth2017, Olivares2019}. 
The $\tt BHAC$ code solves the ideal GRMHD equations with fixed spacetime in geometric units ($GM = c = 1$ and $1/\sqrt{4\pi}$ is included in the magnetic field).

Metric $g_{\mu\nu}$ is implemented in spherical Modified Kerr-Schild coordinates (detail see \cite{Porth2017}). The four-magnetic field are defined as:
\begin{equation}
    b^t = \frac{\gamma\left(B^i v_{i}\right)}{\alpha}, \,\,\,b^{i}=\frac{B^i +\alpha b^t u^i}{\gamma},
\end{equation}
where $\gamma$ is the Lorentz factor, $B^i$ is the Eulerian magnetic field, $v^i$ is fluid three-velocity, $u^\mu$ is fluid four-velocity, and $\alpha$ is the lapse function. Plasma beta, $\beta$, and magnetization, $\sigma$, in this article are expressed respectively as $\beta = p/b^2$ and $\sigma = b^2/\rho$, where the definition of $b^2$ is as follows \citep{Porth2017}. 
\begin{equation}
    b^2 = \frac{B^2+\alpha^2(b^t)^2}{\gamma^2} = \frac{B^2}{\gamma^2}+\left(B^i v_i\right)^2,
\end{equation}
where $B^2 = B^i B_i$. 


The simulations are initialized with a Fishbone-Moncrief hydrodynamic equilibrium torus \citep{1976ApJ...207..962F} with parameters $r_{\rm{in}} = 20 \,r_{\rm g}$ and $r_{\rm max} = 40\, r_{\rm g}$ where $r_{\rm g}\equiv GM/c^2$ is the gravitational radius of the black hole and $M$ is its mass. 
An ideal-gas equation of state with a constant adiabatic index of $\Gamma_{\rm g} = 4/3$ is used \citep{Rezzolla2013}. 

We consider a pure poloidal magnetic field in all cases. By setting up the initial vector potential, the magnetic field configuration is defined as follows:
\begin{equation}
    A_{\rm \phi} \propto \left(\rho-0.01\right)\left(r/r_{\rm in}\right)^3\sin^3\theta\exp\left(-r/400\right).
    \label{Eq: MAD_A}
\end{equation}
To adjust the number of the magnetic loops $N$ on the vertical and radial directions, we multiple $A_{\rm \phi}$ by $\cos\left(\left(N-1\right)\theta\right)$ and $\sin\left(2\pi\left(r-r_{\rm in}\right)/\lambda_{\rm r}\right)$, where $\lambda_{\rm r}$ is the radial loop wavelength and $N$ \footnote{The torus has an opening angle approximately ranging from $-\pi/4$ to $\pi/4$. In $\theta$ direction, $A_{\rm \phi} \propto \cos\left((N-1)\theta\right)$. $N=3$ ensures the polarity along $\theta$ direction does not change and forms only one loop in $\theta$ direction inside the torus. } is the number of magnetic loops in the $\theta$ direction. For all cases, we set $N = 3$. The magnetic field is normalized by setting the minimum value of plasma beta, $\beta \equiv p_{\rm g}/p_{\rm mag}$, where $p_{\rm g}$ is gas pressure and $p_{\rm mag}=b^2/2$ is magnetic pressure. The definition of $b^2$ is calculated by $b^2=b_\mu b^\mu$, where $b^\mu$ is a 4-magnetic field. In our simulations, we use $\beta_{\rm min} = 100$, which is a relatively high value compared with other studies \citep{Chashkina2021, Ripperda2021}. 

In order to excite the magneto-rotational instability (MRI) inside the torus, $2\% $ of a random perturbation is applied to the gas pressure in the torus. 
%
%

The initial parameters of each simulation are listed in Table~\ref{Table: MAD}. Four different sets of parameters are used, the cases with a label ending with "a" means high black hole spin with alternating polarity configuration, "b" indicates the cases of high black hole spin with the same polarity configuration, "c" and "d" represent alternating and same polarity configuration in a non-spinning black hole. Case $\tt M0$ is a reference case of a single loop magnetic field in the torus that forms a typical MAD accretion flow.
Most of the simulations use a grid resolution of $1024\times512$ pixels in radial and azimuthal directions, only the case of $\tt{M80a}$ has been performed with a high resolution of $4096\times2048$ pixels. Our purpose for the higher-resolution simulation is to verify the influence of different resolutions. The comparison of mass-accretion rate and density distribution between the two different numerical resolution cases are presented in the Appendix (Fig.~\ref{fig: 2D_high_low_res}). It is shown that the two different resolution cases have a good agreement. Thus, we conclude that $1024\times512$ grid resolution is enough to study the overall behavior of the simulations. The following discussions are based on the low-resolution simulation results.  

At the inner/outer radial boundaries, we apply standard inflow/outflow boundary conditions by enforcing a copy of the physical variables. At the polar boundaries, reflecting boundary conditions are applied (i.e., $B^\theta \to -B^\theta$, $v^\theta \to -v^\theta$, and other quantities are symmetric). For the azimuthal direction, periodic boundary conditions are employed.

Following our previous work by \cite{Mizuno2021}, we solve the electron thermodynamics during the evolution of single-MHD fluid separately. Based on the two-temperature description, we can obtain the ion ($T_{\rm i}$) and electron temperatures ($T_{\rm e}$) independently from the gas temperature, $T_{\rm gas} = T_{\rm i} + T_{\rm e}$. Here, we neglect the energy exchange rate due to Coulomb coupling, anisotropic thermal heat flux, and radiative cooling which have been considered in previous works \cite[e.g.,][]{Ressler2015,Chael2018,Dihingia2023}.
The grid-scaled dissipation gives the electron heating. This work considers the electron heating mechanism by turbulence and magnetic reconnection. The turbulent heating model is adopted from the simulation of damping of MHD turbulence by \cite{Kawazura2019}, and the heating fraction is given by
 \begin{equation}
     f_{\rm e} = \frac{1}{1+Q_{\rm i}/Q_{\rm e}},\label{eq2}
 \end{equation}
 where
 \begin{equation}
     \frac{Q_{\rm i}}{Q_{\rm e}} = \frac{35}{1 + \left(\beta/15\right)^{-1.4}\exp\left(-0.1 T_{\rm e}/T_{\rm i}\right)}.
 \end{equation}
For the prescription of magnetic reconnection heating, we adopt the fitting function measured in particle-in-cell simulations of magnetic reconnection by \cite{Rowan2017}:
 \begin{equation}
     f_{\rm e} = \frac{1}{2} \exp\left[\frac{-(1-\beta/\beta_{\rm max})}{0.8+\sigma_{\rm h}^{0.5}}\right],
     \label{Eq: reconnection heating}
 \end{equation}
 where $\beta_{\rm max} = 1/4\sigma_{\rm h}$, $\sigma_{\rm h} = b^2/\rho h$ is magnetization as defined with respect to the fluid specific enthalpy $h = 1 + \Gamma_{\rm g} p_{\rm g}/(\Gamma_{\rm g}-1)$. 

 \begin{figure*}
     \centering
          \includegraphics[height=0.4\linewidth]{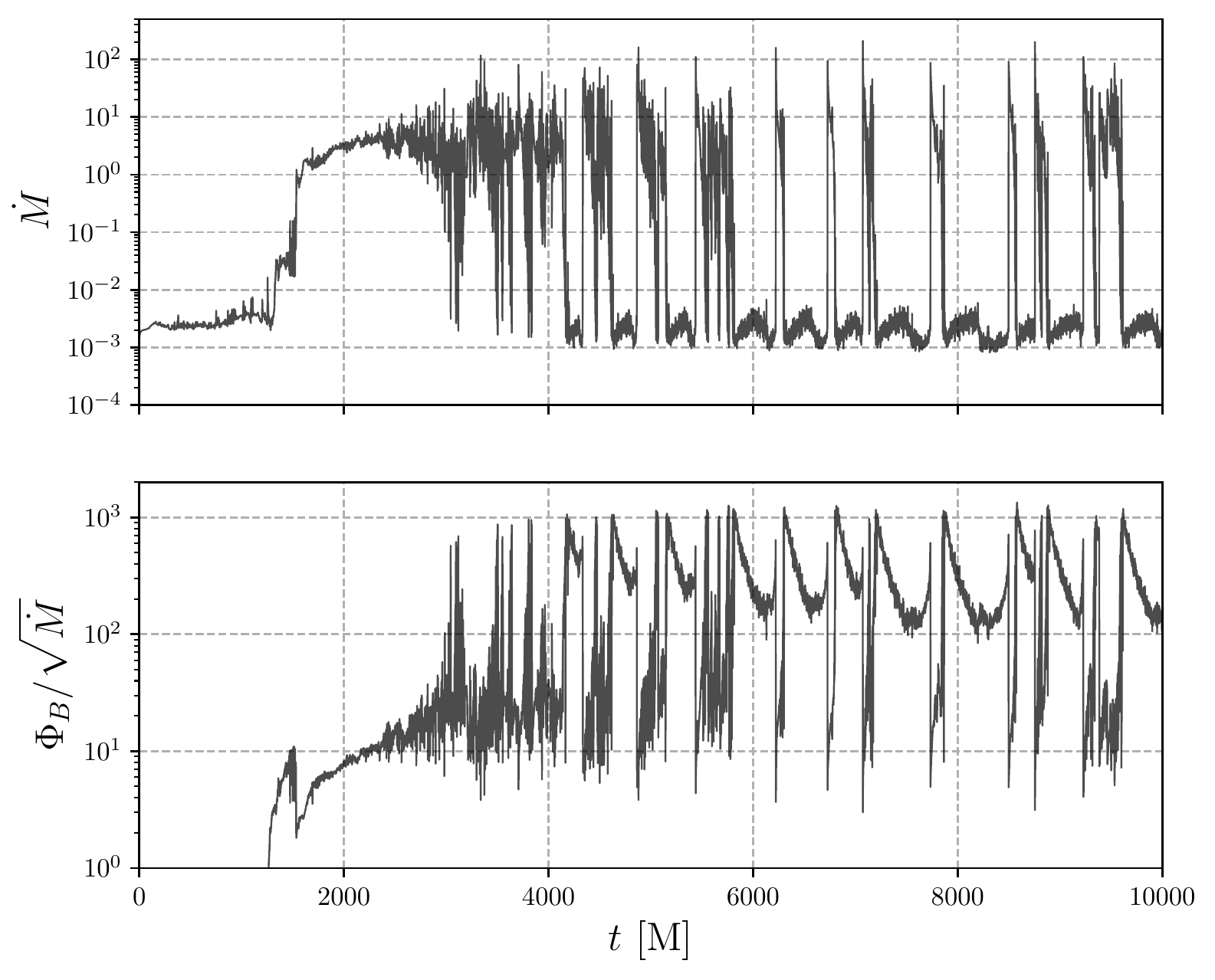}
          \includegraphics[height=0.4\linewidth]{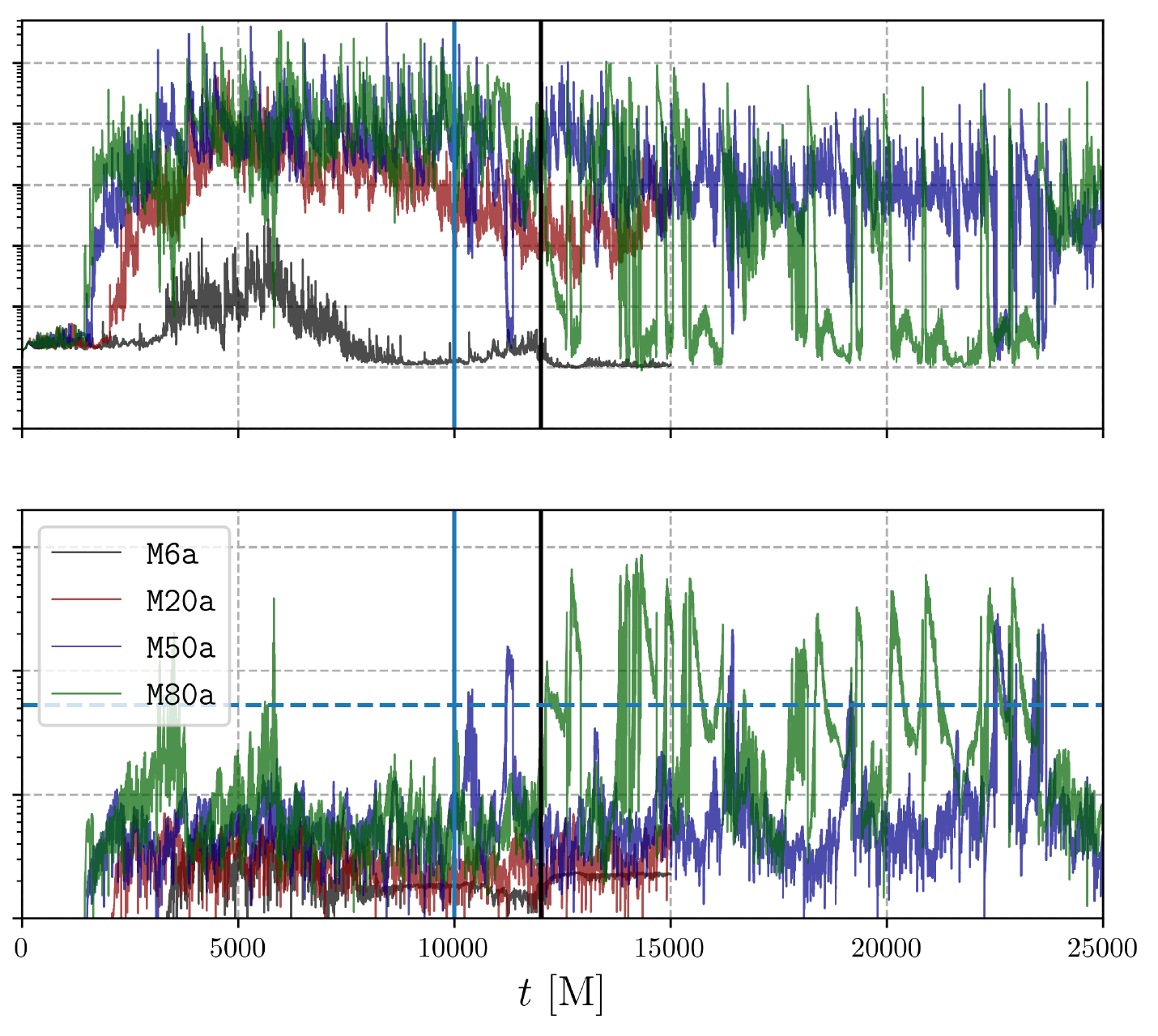}
     \caption{Time evolution of mass-accretion rate and magnetic flux onto the event horizon for a $a = 0.94$ black hole with ({\it left}) the single loop cases, $\tt M0$ and ({\it right}) with alternating polarity magnetic loop cases in different wavelengths, $\tt M6a$ ($\lambda_{\rm r}=6$, black), $\tt M20a$ ($\lambda_{\rm r}=20$, red), $\tt M50a$ ($\lambda_{\rm r}=50$, blue) and $\tt M80a$ ($\lambda_{\rm r}=80$, green), respectively. Blue and black vertical lines represent the transition moment from MAD to SANE of case {\tt M50a} and {\tt M80a} respectively. Horizontal blue dashed line represents the criteria for MAD. Large loop cases reach a quasi-steady state slower, therefore we simulated longer time for these cases.}
     \label{fig: high_spin_alternating}
 \end{figure*}
%
Using different electron heating prescriptions (turbulent and magnetic reconnection heating) we are able to calculate the electron temperature from the sub-grid heating model which is proposed by \cite{Ressler2015}. Since the electron thermodynamics are not affected by the global flow dynamics in our simulations, we evaluate the electron entropy with turbulent \citep[e.g.,][]{Kawazura2019} and magnetic reconnection \citep[e.g.,][]{Rowan2017} prescriptions simultaneously.

\section{Plasma dynamics with different initial magnetic configuration}


 Following the definition of \cite{Porth2019} we calculate the mass-accretion rate measured at the event horizon as 
\begin{equation}
    \dot M = \int_0^{2\pi}\int_0^{\pi} \rho u^r \sqrt{-g}d\theta d\phi, \label{Eq: a_rate}
\end{equation}
And the dimensionless magnetic flux rate measured at the event horizon as
\begin{equation}
    \Phi_{\rm B} = \frac{1}{2}\int_0^{2\pi}\int_{0}^{\pi}\left|B^r\right|\sqrt{-g}d\theta d\phi. \label{Eq: B-flux}
\end{equation}

We should note that in units employed here, the dimensionless magnetic flux rate at the black hole horizon differs from the definition in \citet{tchekhovskoy_efficient_2011} by a factor of $\sqrt{4 \pi}$.

The left panel of Fig.~\ref{fig: high_spin_alternating} represents the time evolution of mass-accretion rate $\dot M$ and magnetic flux $\Phi_{\rm B}$ of the single large poloidal magnetic field loop case $\tt M0$. In this case, without adding any small-scale magnetic loops in the torus, the simulation is reached a typical MAD state of accretion flows. 
As shown in the left panel of Fig.~\ref{fig: high_spin_alternating}, there are clear periodic patterns of the suppression (termination) of the accretion flows which is the signature of the MAD state. 
By comparing with this single loop case, we investigate the influence of different multiple loop configurations.

\subsection{Multiple magnetic loops with alternating polarity}
\begin{figure}
    \centering
         \includegraphics[width=\linewidth]{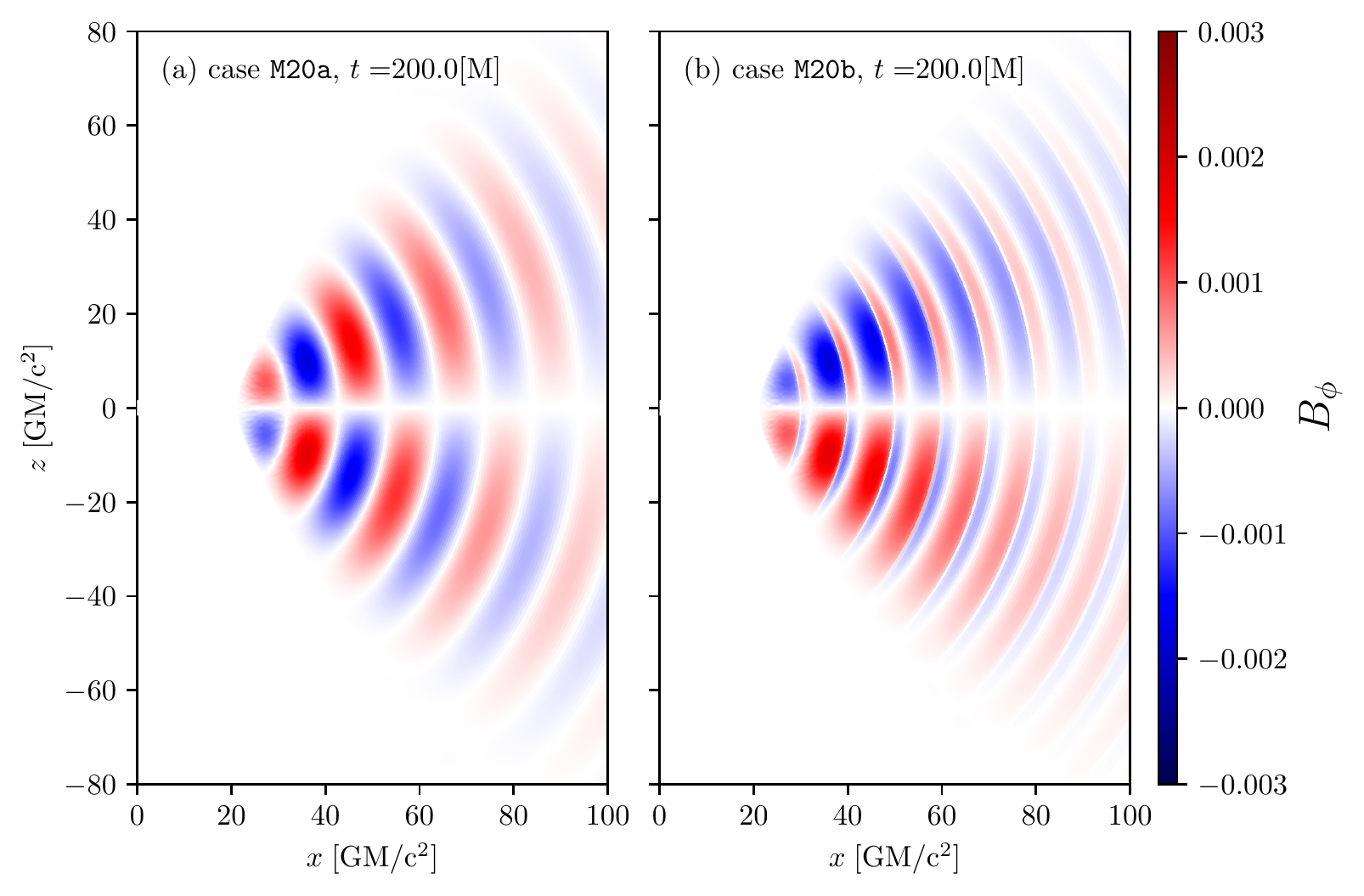}
    \caption{Distribution of $B_{\rm \phi}$ amplified by differential rotation of torus at the beginning of the simulations. {\it Left} is alternating polarity configuration case ($\tt M20a$), {\it right} is same polarity configuration case ($\tt M20b$).}
    \label{fig: M80_B}
\end{figure}
\begin{figure}
    \centering
	\includegraphics[height=1.\columnwidth]{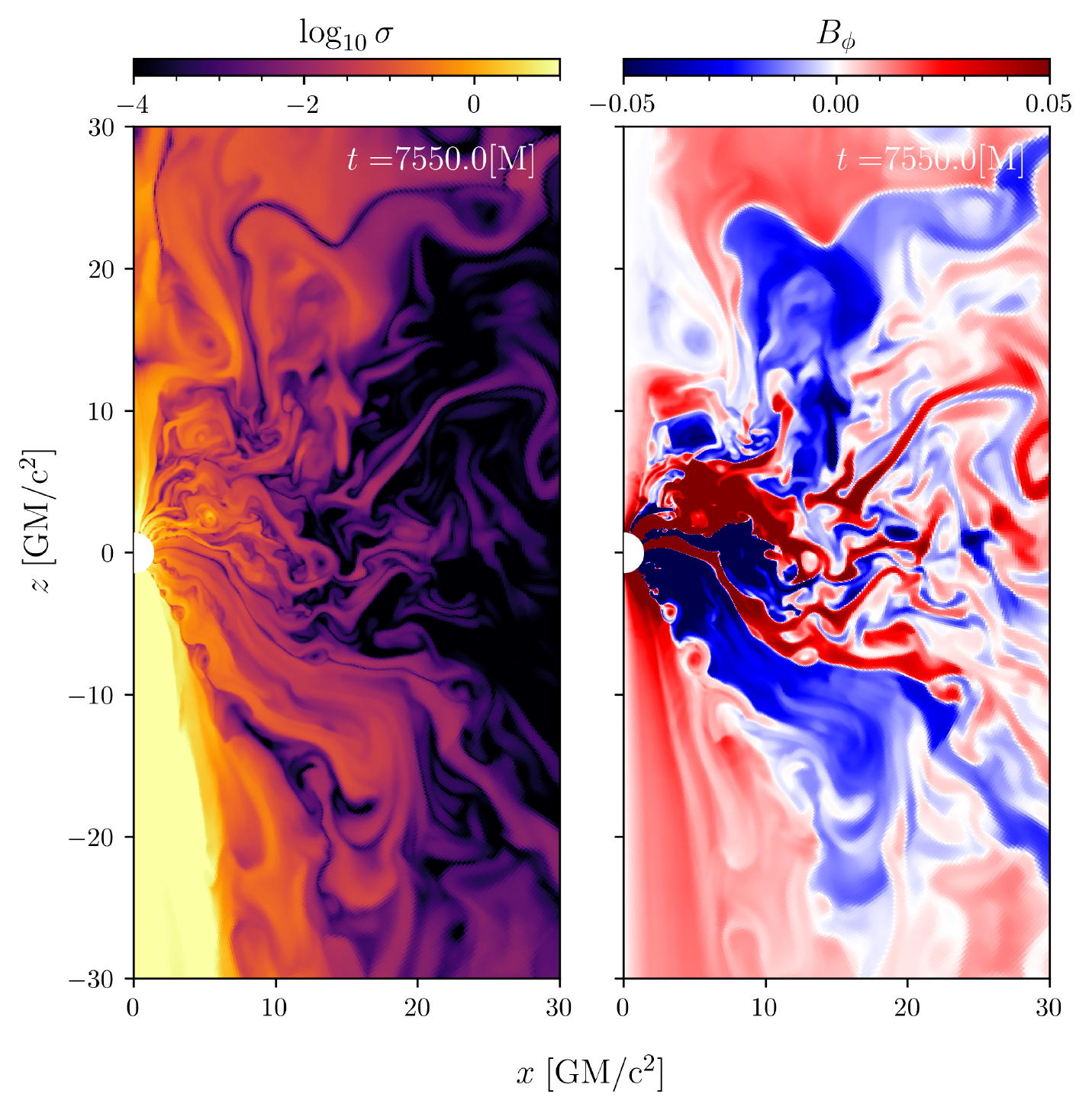}
    \caption{The snapshot of magnetization $\sigma$ distribution ({\it left}) and toroidal magnetic field $B_{\rm \phi}$ distribution ({\it right}) for $\tt M20a$ at $t=7550\; \rm M$. The upper part of the jet disappears due to the highly chaotic accretion flow.}
    \label{fig: M20a_2D_magnetization}
\end{figure}

\begin{figure*}
    \centering
         \includegraphics[width=0.8\linewidth]{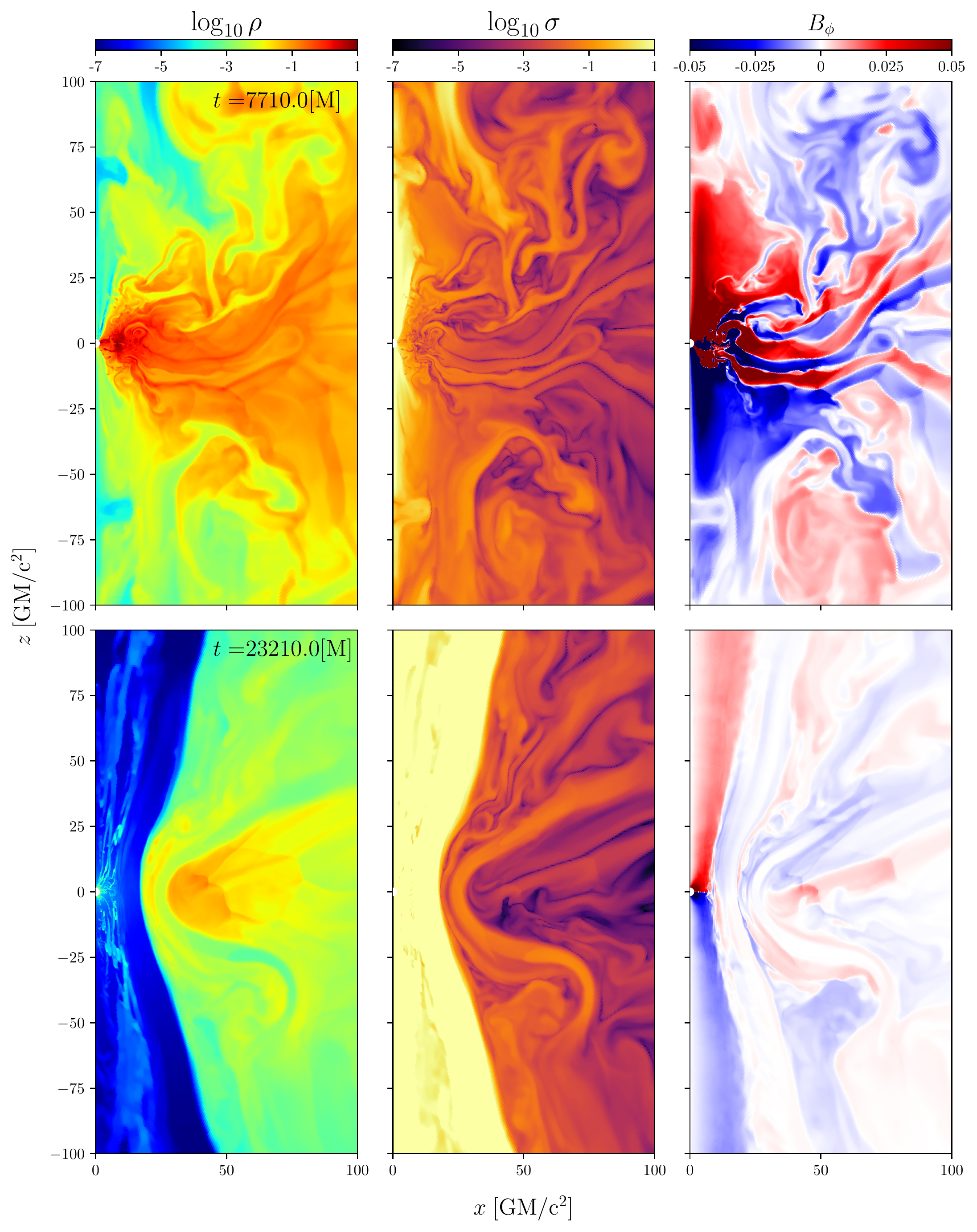}
         \caption{Snapshot images of the logarithmic density ($\rho$, {\it left}), magnetization ($\sigma$, {\it middle}), and toroidal magnetic field ($B_{\rm \phi}$, {\it right}) for the alternating polarity magnetic loop case with $\lambda_{\rm r}=80$ ($\tt M80a$) during SANE ($t=7710$ M, {\it upper}) and MAD ($t=23210$ M , {\it lower}) phases, respectively.}
    \label{fig: 2D_M80a}
\end{figure*}

%
The right panel of Fig.~\ref{fig: high_spin_alternating} shows the time evolution of accretion rate $\dot M$ and magnetic flux $\Phi_{\rm B}$ for the cases of highly spinning black hole ($a = 0.9375$) with alternating polarity loops. 
Different colored lines indicate the cases with the different wavelengths (Cases $\tt M6a$, $\tt M20a$, $\tt M50a$, and $\tt M80a$ with loop wavelength $\lambda_{\rm r}=6$, $20$, $50$, and $80$ respectively).


For case $\tt M6a$, the torus is initiated with small magnetic loops (wavelength $\lambda_{\rm r}=6$), compared with the torus size. 
Thus, before MRI is developed enough, reconnection occurs inside the torus and most of the magnetic energy is dissipated, which gives rise to the comparatively lower accretion rate and magnetic flux seen in the right panel of Fig.~\ref{fig: high_spin_alternating}. 

Although we have enough numerical resolution to capture MRI in the torus (MRI quality factor $Q_{\rm r} \gg 10$), MRI becomes weaker due to the dissipation between each small loop.

Cases $\tt M20a$ and $\tt M50a$ show a similar accretion rate and magnetic flux evolution at the early stage with the $\tt M0$ case. 
The important difference is whether they transit to MAD state or not in later times. 
MRI is well-sustained in long-time evolution due to large initial torus size.
Case $\tt M20a$ does not reach the full MAD state because the signature of the MAD state which is the oscillatory drop of mass accretion rate and magnetic flux is not seen clearly (more discussion is in Sec.3.3). Although we observe some magnetic field accumulation and accretion rate decrease during the first two loops fall onto the black hole from $2000$ to $4000\, \rm M$. 
As the accretion process continues, the substantial reduction in the flux evolution becomes less obvious. This means that the MAD state is not fully developed. 

Figure~\ref{fig: M20a_2D_magnetization} shows magnetization and toroidal magnetic field distribution at $t=7550\,{\rm M}$ of case {\tt M20a}. The high magnetization region at the lower component of the funnel region shows a one-sided jet structure. The lower region close to the horizon shows a current sheet that leads the formation of plasmoids generated by the changing polarity (see the {\it right} panel in Fig.~\ref{fig: M20a_2D_magnetization}). Similar structures are seen in previous work by \cite{Nathanail2020}. The details of the plasmoid formation mechanism will be investigated in the next section. 

For case $\tt M50a$, due to the bigger size of the magnetic loops, the dissipation inside the torus becomes weaker. The MRI is well-developed and sustained long-time, which leads to a higher mass accretion rate $\dot M\sim 10$. The magnetic field accumulation on the horizon becomes stronger. Accretion continues as in the SANE state during half of the simulation time. The saw-tooth behavior in the flux evolution starts to observe since $t\approx 11500 \, \rm M$, that the accretion flow is fully quenched by the strong magnetic field on the horizon.

In the case of $\tt M80a$, the situation is similar to model $\tt M50a$. The accretion rate in these cases is close to the typical single loop MAD case seen in the left panel of Fig.~\ref{fig: high_spin_alternating}. The upper panels of Fig.~\ref{fig: 2D_M80a} show the distribution of physical quantities which represents the SANE state at early simulation time ($t=7710\,M$). Due to the accretion of multiple alternating polarity loops, the funnel region is filled with plasma with higher density than in typical single-loop cases.
The bigger magnetic loops contribute to less magnetic dissipation inside the torus. Thus, the magnetic field is able to accumulate on the black hole horizon which causes the suppression of accretion flows, i.e. entering the MAD state. 
After $t=12000\,\rm M$, the magnetic field is strong enough to periodically suppress the accretion flow, similar to the single loop case $\tt M0$. However, by comparing the left panel of Fig.~\ref{fig: high_spin_alternating} with the right panel of it, the accretion rate and magnetic flux evolution in case $\tt M80a$ are different from the case $\tt M0$, due to the effect of changing polarity in the funnel region. 
At later times, the MAD-like behavior is clearly represented (see lower panels of Fig.~\ref{fig: 2D_M80a}). 
%
\subsection{Multiple magnetic loops with same polarity}\label{sec: GRMHD_sam}
\begin{figure}
	\centering
	\includegraphics[height=.8\columnwidth]{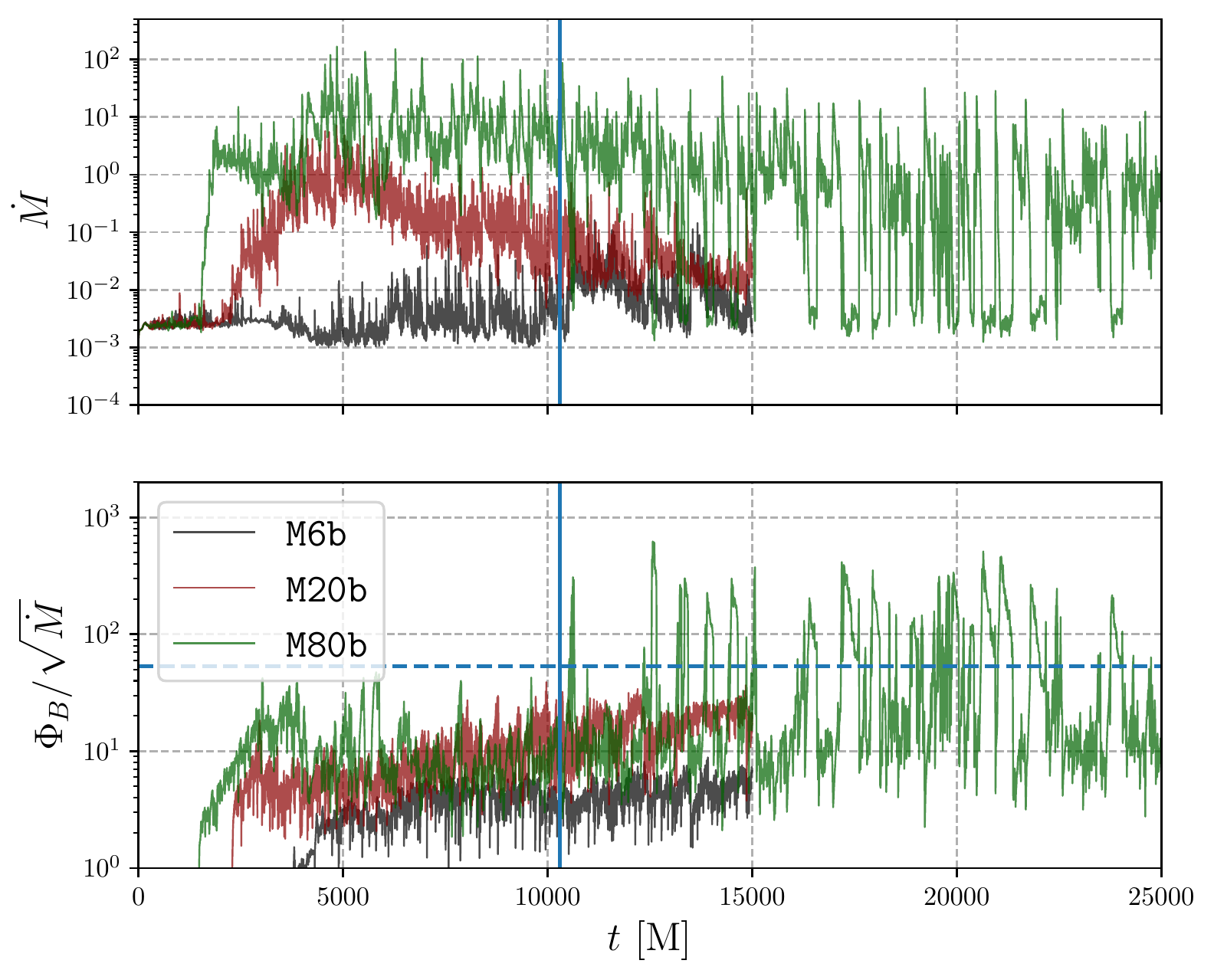}
    \caption{Same as Fig.\ref{fig: high_spin_alternating} but shown for the spinning black hole cases with multiple magnetic loops with the same polarities. Different colors indicate different loop wavelengths, $\tt M6b$ ($\lambda_{\rm r}=6$, black), $\tt M20b$ ($\lambda_{\rm r}=20$, red), and $\tt M80b$ ($\lambda_{\rm r}=80$, green), respectively. Horizontal blue dashed line marks the criteria of MAD. The vertical blue line is the moment of SANE to MAD transition of case {\tt M80b}. Due to the slower evolution of case {\tt M80b}, we extended the simulation of this case to $25000\,\rm M$.}
    \label{fig:Mdot_high_sam}
\end{figure}


Figure~\ref{fig:Mdot_high_sam} shows the mass accretion rate and the magnetic flux accumulation rate on the horizon for the spinning black hole cases with multiple magnetic loops of the same polarity. Different colors indicate different wavelengths, $\tt M6b$ ($\lambda_{\rm r}=6$, black), $\tt M20b$ ($\lambda_{\rm r}=20$, red), and $\tt M80b$ ($\lambda_{\rm r}=80$, green), respectively. In comparison with alternating polarity cases seen in the right panel of Fig.~\ref{fig: high_spin_alternating}, the differences in the mass accretion rate and the magnetic flux rate are large and reflect the difference in the time evolution of the magnetic loops inside the torus.

The case $\tt M6b$ has relatively small magnetic loops as the initial condition. Consequently, there is much dissipation inside the loops at the early time. 
In Fig.~\ref{fig:Mdot_high_sam}, from $2000 \, \rm M$ to $6000 \, \rm M$, the accretion rate of case $\tt M6b$ is very low.
During that time period, MRI is not well developed due to the weak magnetic field. A similar trend is also seen in the case of alternating polarity with small loop wavelength (see the right panel of Fig.~\ref{fig: high_spin_alternating}). 
After magnetic reconnection in the loops has finished, there is almost one weak single loop is left. 
MRI starts to develop and accretion proceeds at a later time. Magnetic flux starts to accumulate on the event horizon from later time $\sim t=5000\,M$ (see also Fig.~\ref{fig:Mdot_high_sam}).

The situation in the $\tt M20b$ case is similar to the $\tt M6b$ case. Because the magnetic loop scale becomes larger which includes larger differential rotation, the deference of polarities in loops becomes stronger. Dissipation by magnetic reconnection inside the torus is partially working and finished earlier. Then, MRI starts to grow quickly which makes a larger scale turbulent structure than that seen in the $\tt M6b$ case (see Fig.~\ref{fig:Mdot_high_sam}). It forms a typical SANE accretion flow. As magnetic amplification proceeds, $\Phi_{\rm B}$ monotonically increases. However, we can not see clear evidence of the transition to the MAD state during the simulation period.

As seen in the alternating polarity case, when the magnetic loop becomes large, the dissipation by magnetic reconnection inside the torus becomes weak. 
In the $\tt M80b$ case, The time evolution in the early stage is similar to that seen in the single loop case. MRI produces large-scale turbulent field structures. 
Before $t = 10000\,\rm M$, the accretion flow is basically a SANE state with continuous magnetic flux accumulation near the horizon. Once enough magnetic flux is accumulated on the horizon, accretion flows transit to the MAD state. We see MAD-like behavior in later simulation time ($t > 10000\,\rm M$, see Fig.~\ref{fig:Mdot_high_sam}).

\subsection{Transition from SANE to MAD regime}
\begin{figure}
    \centering
         \includegraphics[width=\linewidth]{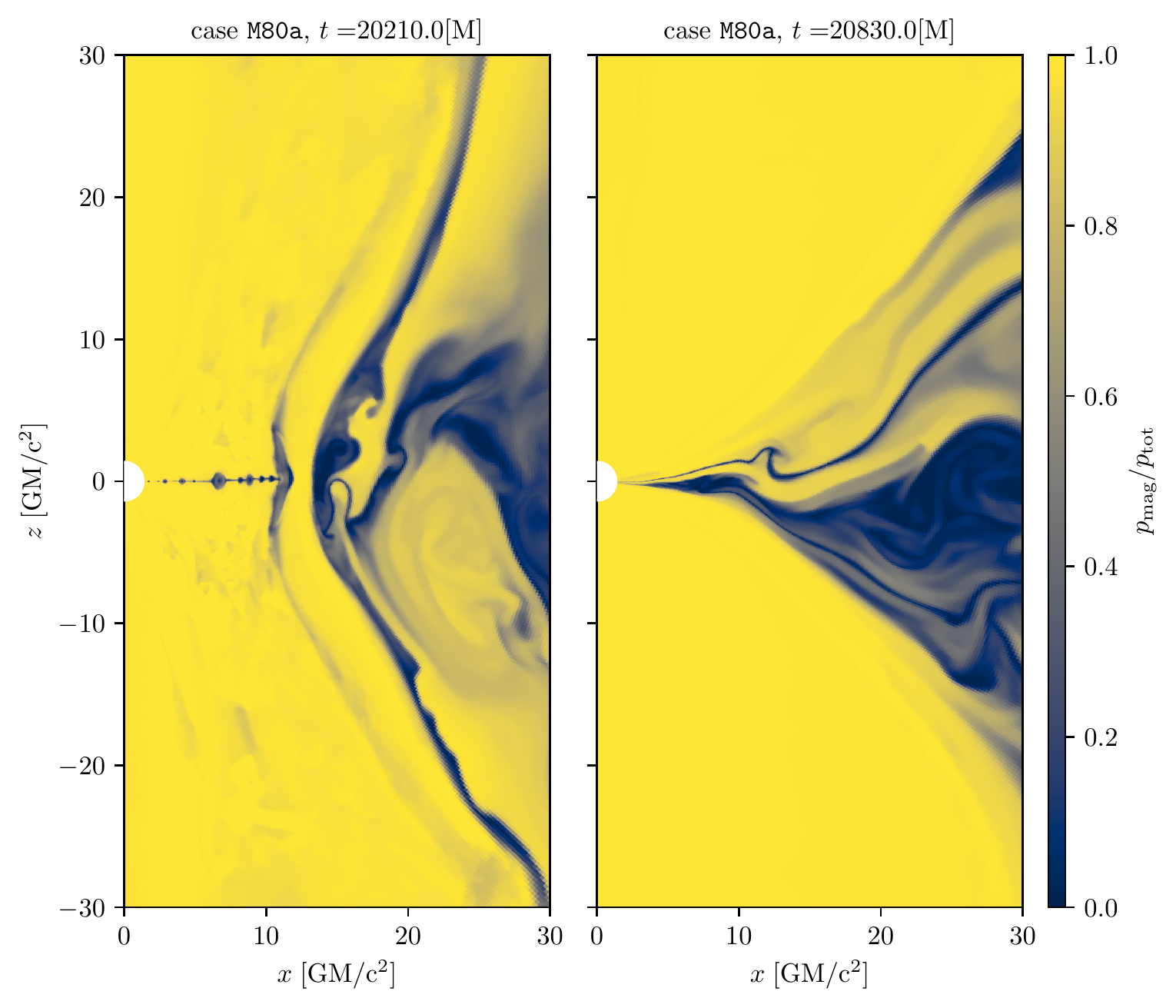}
    \caption{The ratio of magnetic pressure and total pressure when the accretion disk is fully arrested by the magnetic field (MAD) ({\it left}) and after MAD ({\it right}).}
    \label{fig: B_phi_ratio}
\end{figure}
\begin{figure}
    \centering
         \includegraphics[width=\linewidth]{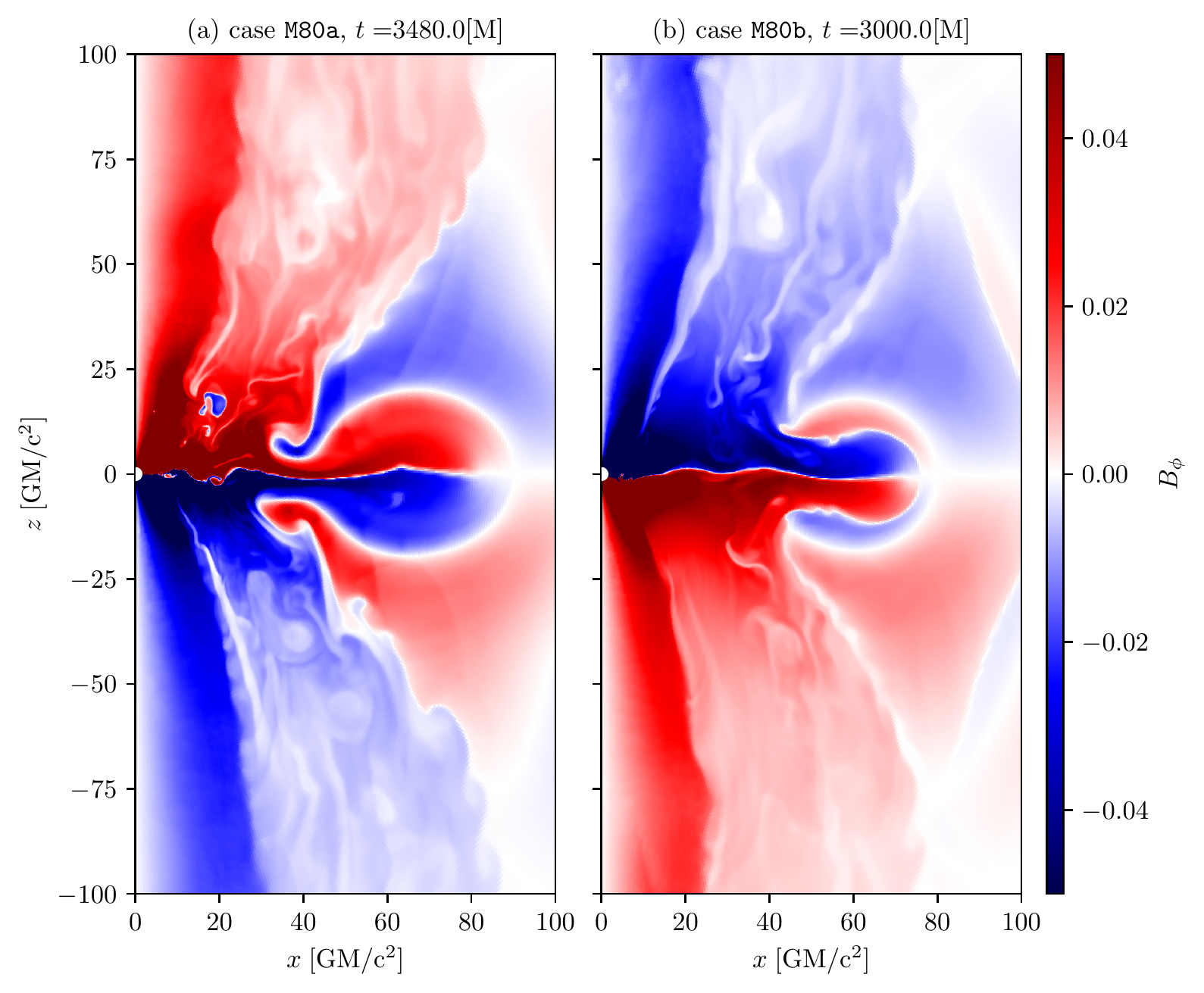}
    \caption{Distribution of toroidal magnetic field $B_{\rm \phi}$ for the alternating polarity multi-loop case $\tt M80a$ at $t=3480  \, \rm M$ ({\it left}) and for the same polarity multi-loop case $\tt M80b$ at $t=3000  \, \rm M$({\it right}). 
    }
    \label{fig: M80_RT_instability}
\end{figure}

%

In previous 2D GRMHD simulations \citep[e.g.,][]{tchekhovskoy_efficient_2011}, the criterion to enter the MAD state is that the normalized magnetic flux on the horizon becomes higher than the threshold value of $\sim 15$. In our 2D simulations, this criterion is applicable although it differs by a factor of $\sqrt{4 \pi}$ due to units employed, $\sim 53$. 
In the MAD state, the mass accretion rate drops by more than 3 orders of magnitude. This behavior forms a saw-tooth structure in the evolution of the mass accretion rate (see Fig.~\ref{fig: high_spin_alternating}). Therefore, the criteria for entering the MAD state in this work is the disruption of the accretion process.
Figure~\ref{fig: 2D_M80a} shows the distributions of density, magnetization, and toroidal magnetic field in the SANE ({\it upper panels}) and MAD states ({\it lower panels}) of the case $\tt M80a$.
The magnetic field on the horizon is continuously accumulated if the magnetic field that is dragged through accretion has the same polarity. At the same time, the toroidal magnetic field is amplified due to the frame-dragging effect of a rotating black hole. 
At the MAD state, magnetic pressure is largely dominant near the horizon. As seen in Fig.~\ref{fig: B_phi_ratio}, the ratio of $p_{\rm mag}/p_{\rm tot}$ is nearly 1 \citep{2022ApJ...939...31M}, where $p_{\rm tot} = p_{\rm mag}+p_{\rm gas}$. Such strong magnetic pressure resists the mass accretion onto the black hole and makes the MAD state.

Although both same and alternating polarity cases show a similar trend of transition from SANE to MAD, especially for the simulations with large magnetic loops after $t=10000\,\rm M$
 (see Figs.~\ref{fig: high_spin_alternating} and \ref{fig:Mdot_high_sam}), which shows the normalized magnetic flux exceeds the threshold value of MAD state. The dynamics in the two situations are quite different. 

As discussed above, in the same polarity cases, the dissipation of the magnetic field via magnetic reconnection in the torus slows down the accretion onto the black hole and the magnetic field accumulation on the event horizon in early simulation time. If a strong enough magnetic field is left over after the initial magnetic dissipation inside the torus, accumulated $\Phi_{\rm B}$ is large enough to make MAD state at a later time. 

On the other hand, in the alternating polarity cases, there is no dominant polarity in the torus. The polarity of each magnetic loop alternates along the radial direction. 
%
Reconnection between different loops largely increases the complexity of the magnetic field distribution inside the torus. 
A large amount of small-scale magnetic field with different polarities is produced during these periods (see the {\it right} panel of Fig.~\ref{fig: M20a_2D_magnetization}).

In the left panel of Fig.~\ref{fig: M80_RT_instability}, the second loop with opposite polarity of $B_{\rm \phi}$ component (blue in the upper hemisphere and red in the lower hemisphere) accretes onto the horizon via a narrow channel. 
As the magnetic loops with different polarities touch the horizon, the polarity of the jet sheath region is quickly changed. 
Rapid polarity transitions reset the magnetic field accumulation on the horizon. The SANE state is continued until magnetic field accumulation is reached the threshold to the MAD state, which is only seen in the cases with enough large magnetic loops such as the cases of $\tt M50a$ and $\tt M80a$. Thus, in the alternating magnetic loop cases, the size of magnetic loops is a key ingredient in whether it reaches the MAD state.

%

\cite{Chashkina2021} reports accretion flow transits from MAD to SANE in their single loop simulation (model $\tt 2D1$ in their work). They used rather lower $\beta_{\rm min}$ ($=1$) which makes a stronger initial magnetic field.
Although they used low $\beta_{\rm min}$, their initial torus size is also smaller than our cases, leading to a weakened magnetic field. At the late time, it no longer arrests the disk and therefore transits to SANE flow. In our case $\tt M80a/b$, though the basic mechanisms of the two cases are different, both of them are the result of the multiple loop initial magnetic configurations. 

\subsection{Non-rotating black hole cases}

 \begin{figure}
     \centering
 	\includegraphics[height=.8\columnwidth]{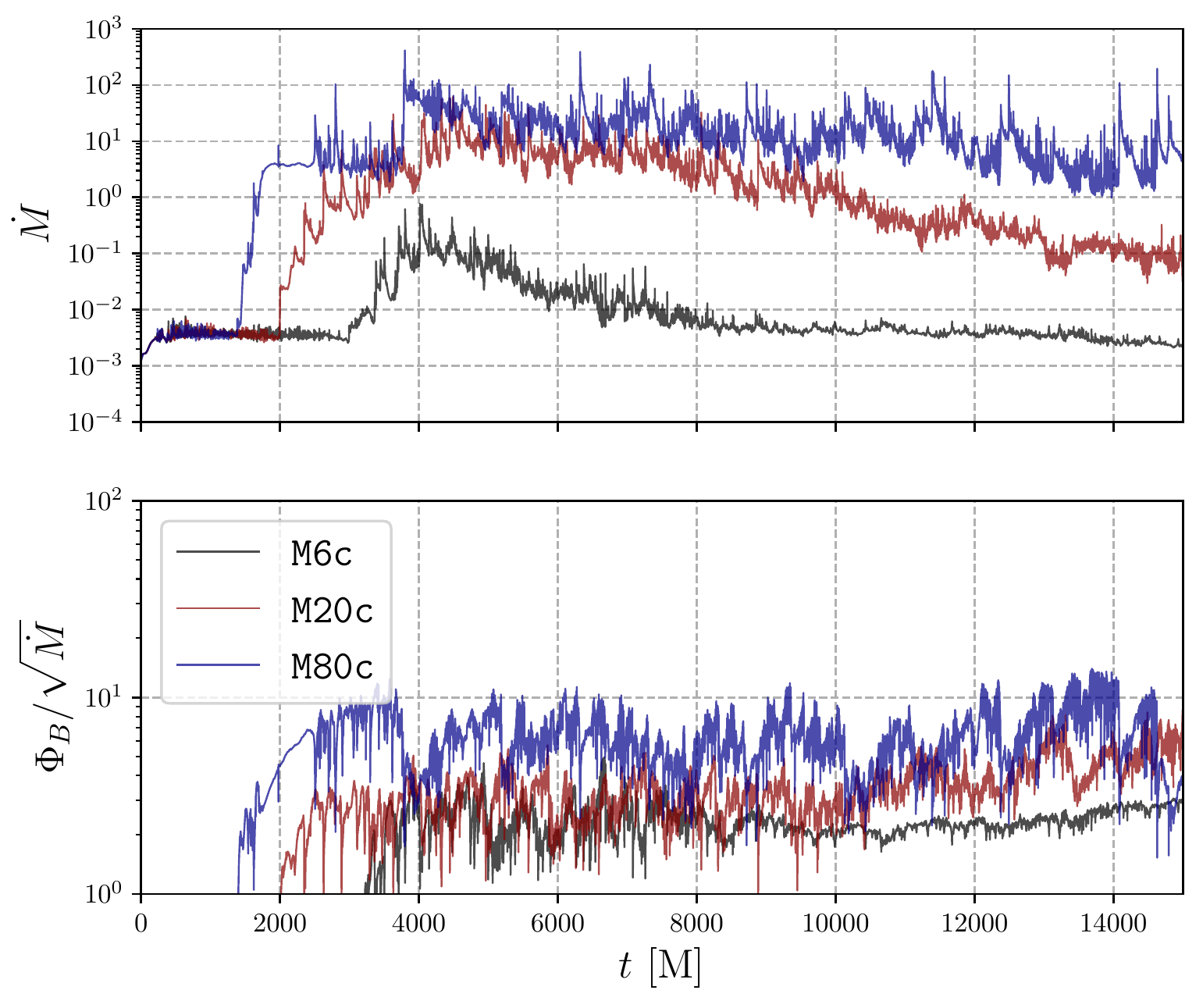}
 	\includegraphics[height=.8\columnwidth]{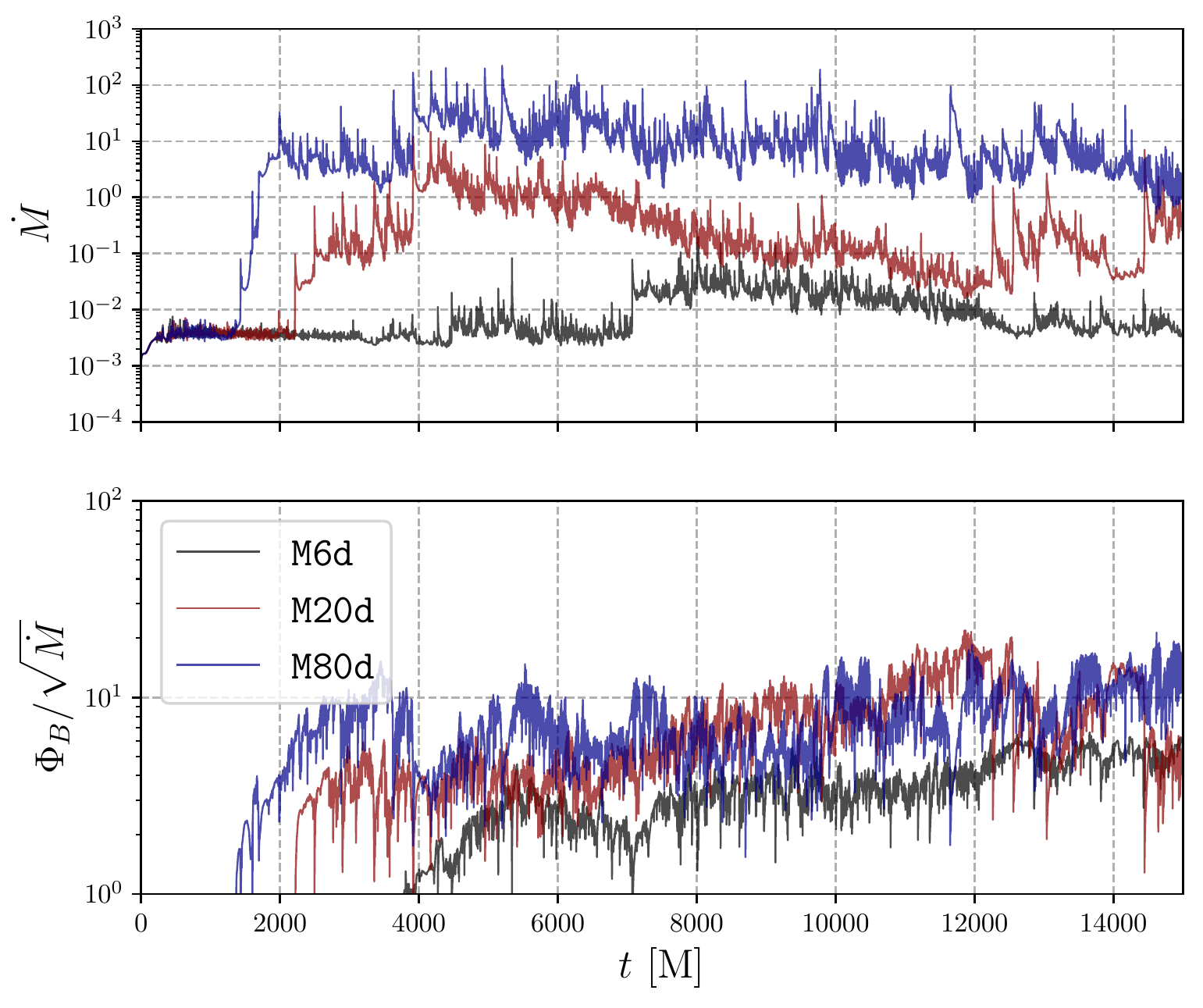}
     \caption{Same as Fig.\ref{fig: high_spin_alternating} but shown for non-spinning black hole cases with multi magnetic loops and alternating polarities ({\it upper panels}) and with same polarities ({\it lower panels}).}
     \label{fig:Mdot_zero_alt}
 \end{figure}

As seen in Fig.~\ref{fig:Mdot_zero_alt}, the time evolution of the mass accretion rate and magnetic flux  for the non-rotating black hole cases are basically the same as those in the rotating black hole cases, except for the large loop wavelength cases $\tt M80c$ and $\tt M80d$.
In general, the black hole spin can not affect the accretion process in the torus much. However, it has a large influence on the jet power and structure \citep{EHTpaperV, Nathanail2021}.
The tori in the shorter wavelength cases are very similar to the ones in the spinning black hole cases. 
Similar to the rotating black hole with alternating polarity cases, the magnetic field on the black hole horizons of case $\tt M6c$ and $\tt M20c$ are experienced rapid polarity changes, which makes them hard to form steady jets, usually one-side transient weak jets and outflows are created. 
On the other hand, the same polarity cases $\tt M6d$ and $\tt M20d$ have weak steady outflows. 
In the simulations, the turbulent magnetic field developed by MRI leads to the dissipation of the magnetic field inside the torus. On the black hole horizon, magnetic field with the same polarity are continuously accumulated which makes relatively high magnetization in the funnel region. Magnetic pressure is the cause of the formation of the jets in these cases.
In the alternating polarity cases, we do not observe such high magnetization in the funnel region due to the accretion of different polarity loops. 

For the cases with longer wavelength magnetic loops, $\tt M80c$ and $\tt M80d$, the dynamics are slightly different from those of the rotating black hole cases. As shown in Fig.~\ref{fig:Mdot_zero_alt}, the time evolution of the mass accretion rate and magnetic flux for $\tt M80c$ and $\tt M80d$ are basically the same as those in the rotating black hole cases with large magnetic loops (see Fig.~\ref{fig: high_spin_alternating} and \ref{fig:Mdot_high_sam}). However, in both polarity cases, the non-rotating ones do not reach the MAD phase.
This implies that the black hole spin is essential for the magnetic field accumulation on the horizon. The frame-dragging effect of a rotating black hole works as additional magnetic field amplification in the black hole which increases the magnetization on the horizon compared with the non-rotating one. This extra magnetic field amplification by frame-dragging effect is essential for the formation of the MAD flow for the multiple loop cases with a longer wavelength.

\section{Plasmoid Formation}


In our simulations, we demonstrate that a shorter wavelength multiple loop configuration produces accretion flows similar to be SANE state, even though a relatively larger torus is considered initially. Turbulence developed by MRI and dissipation by magnetic reconnection inside the torus generates a complicated magnetic field structure. Such chaotic structure leads to sudden polarity changes and one-sided jet formation (see Fig.~\ref{fig: M20a_2D_magnetization}).
%
%
This scenario is also observed in \cite{Nathanail2021}. During the polarity change, the magnetic field with opposite polarities touch each other, and magnetic energy dissipates by magnetic reconnection which leads to the formation of the plasmoids. In addition to magnetic reconnection, we have another channel to make plasmoids through turbulent motion via plasma instability such as Kelvin-Helmholtz (KH) and tearing instabilities \citep{2022ApJ...929...62B}. The relation between plasmoid formation and plasma instabilities has been investigated in previous studies \citep[e.g.,][]{Loureiro2012,Ni2017}. KH instability is one of the channels to form plasmoids at the shear boundary between funnel and sheath regions. Tearing instability is usually associated with magnetic reconnection by an extended thin current sheet which produces plasmoid chains. \cite{Ripperda2021} have discussed the development of plasmoid chains in the equatorial plane in the MAD state. 
Similar plasmoid chain development is also found in the MAD state of the larger loop cases in our simulations.
Here we investigate the properties of the plasmoids in more detail.

\begin{figure}
    \centering
	\includegraphics[width=0.7\columnwidth]{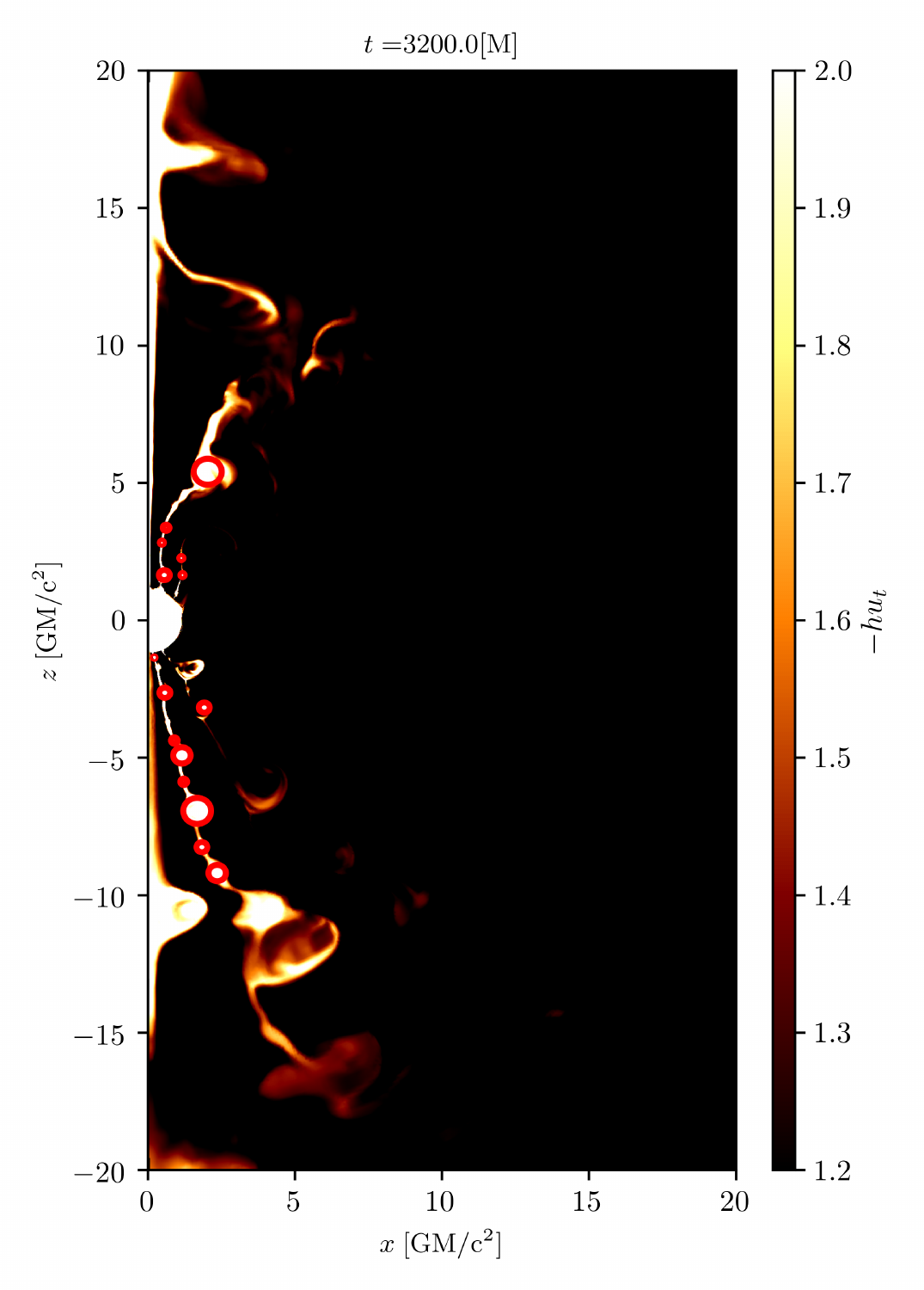}
    \caption{The distribution of the geometric Bernoulli parameter $-hu_{\rm t}$ for the rotating black hole case in alternating polarity loops with $\lambda_r = 50$ ($\tt M50a$) at $t=3200\, \rm M$. Plasmoids are detected and marked with red circles.}
    \label{fig: case_M50a_hut}
\end{figure}

\begin{figure*}
	\includegraphics[width=0.8\textwidth]{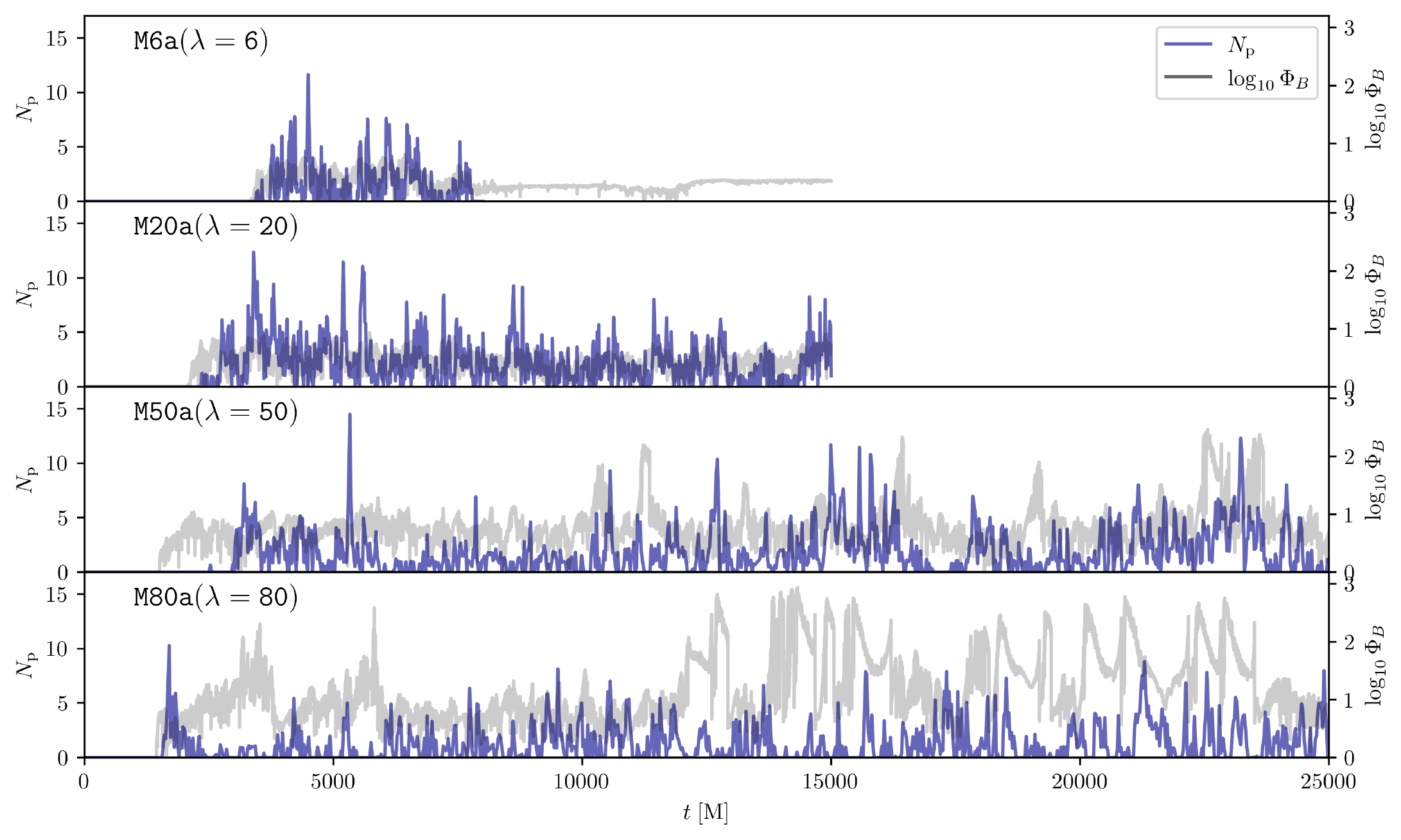}
    \caption{Time evolution of the number of plasmoids for the rotating black hole cases of alternating polarity with different wavelength. Blue lines indicate the plasmoid amount and gray lines represent the magnetic flux rate ($\Phi_{\rm B}$), respectively.}
    \label{fig: plasmoid_amount_alt}
\end{figure*}
\begin{figure*}
	\includegraphics[width=0.8\textwidth]{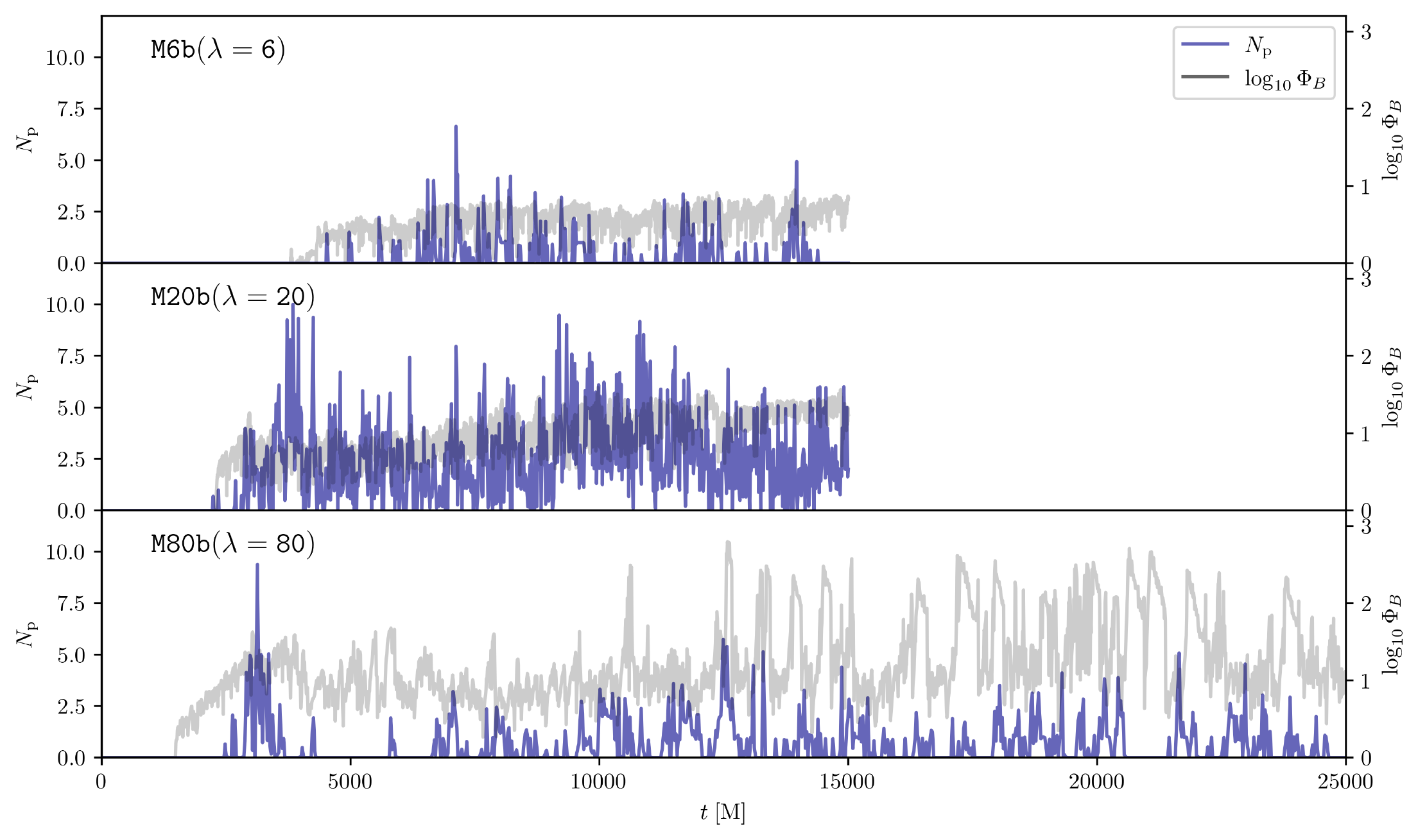}
    \caption{Same as Fig.~\ref{fig: plasmoid_amount_alt} but for the rotating black hole cases with the same polarity magnetic loops.}
    \label{fig: plasmoid_amount_sam}
\end{figure*}

\begin{figure*}
	\includegraphics[height=.6\textwidth]{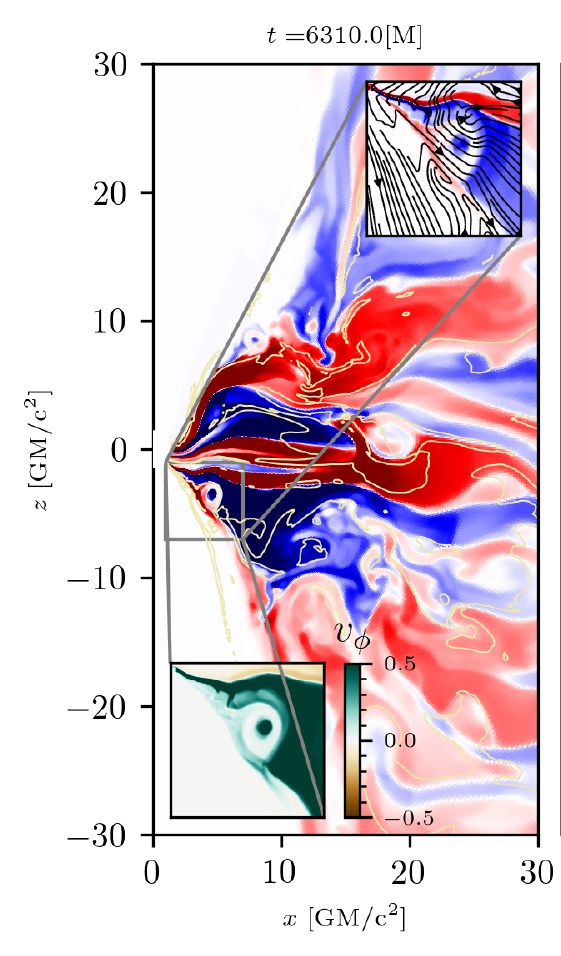}
	\includegraphics[height=.6\textwidth]{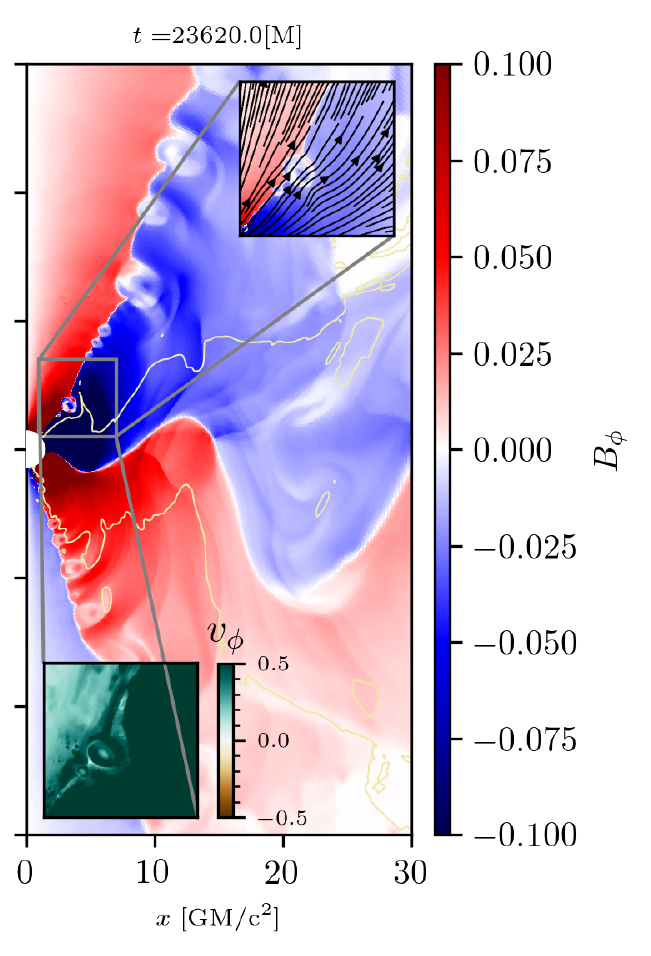}
    \caption{Distribution of toroidal magnetic field at $t=6310\,\rm M$ of case {\tt M80d} ({\it left}) and at $t=23620\,\rm M$ of case {\tt M80a} ({\it right}).
    The zoomed region focused on the plasmoids developed by different mechanisms. The streamlines in the zoomed regions represent the poloidal fluid velocity and the green-brown color indicates toroidal velocity.
    The left panel shows the coexistence of KH instability and reconnection in the formation of plasmoids. The right panel represents the plasmoids from tearing instability in the current sheet.
    Yellow contours in each panel show the transition of gravitationally bounded/unbounded region via the geometric Bernoulli parameter $u_{\rm t}=-1$.
    }
    \label{fig: KHI and reconnection}
\end{figure*}

In order to investigate the properties of each plasmoid, we need to detect them in the simulations.
A plasmoid detection method is proposed in \cite{Nathanail2020}. Following their method, we implement
plasmoid tracking in our simulations using the $\tt scikit-image$ package \citep{VanDerWalt2014}. We capture the position and size of the plasmoids in a box with the range from $0\sim20\, \rm M$ in the radial direction from $-20$ to $20 \, \rm M$ in the axial direction in the simulation snapshots. The snapshots have a time cadence of $10\,\rm M$.
In order to better extract the plasmoids in the jet region alone, we use the Bernoulli constant $-hu_{\rm t}$ and select the region where it is larger than $1.02$ which corresponds to unbounded plasma. Through this selection procedure, we can effectively exclude the torus and most parts of the disk winds with slow outflow velocity. 
Figure~\ref{fig: case_M50a_hut} shows a snapshot image of the Bernoulli constant for the case of $\tt M50a$ at $t=3200\, \rm M$. It is seen that the plasmoids are clearly detected in the jet region. 
We note that in this plasmoid detection method, the plasmoids on the equatorial plane developed by tearing instability are mostly excluded because they are bounded. 
Performing the plasmoid detection, we can obtain the locations and approximate radii of the plasmoids. Due to the limit of numerical resolution, we only count the plasmoids which have a radius larger than $1/30\, \rm M$.
This selection implies that only the large energetic plasmoids will only contribute to the observed signature, such as flares.

The plasmoid formation rate is influenced by the occurrence of reconnection and KH instabilities at the boundary between the jet and sheath region. 
Figures~\ref{fig: plasmoid_amount_alt} and \ref{fig: plasmoid_amount_sam} show the time evolution of the number of plasmoids $N_{\rm p}$ for the rotating black hole cases with alternating and same polarity loops, respectively. The time evolution of the magnetic flux $\Phi_{\rm B}$ accreted onto a horizon is overlaid in these figures as light gray solid lines. 
In general, rotating black hole cases produce powerful jets even initialized with different magnetic field configurations.
Such powerful jets are good for the formation of plasmoids. Because the high-velocity outflow in the funnel region and inflow in the sheath region generates strong velocity shear at the boundary that leads to the growth of KH instability which increases the plasmoid formation rate. 
Due to the multi-loop configuration, there are strong currents at the sheath region.
The magnetic reconnection and the tearing instabilities at the current sheet in this region would be the main contributors to the formation of plasmoids.



From the time evolution of magnetized accretion flows, we see the transition from SANE to MAD state in the larger loop cases in the simulations with alternating polarity loops. On the other hand, in smaller loop cases, only the SANE state turns out. 
In the small loop cases ($\tt M6a$ and $\tt M20a$), the initially ordered loops mix up shortly after the simulation starts. For example, in the case $\tt M20a$, as shown in the second panel of Fig.~\ref{fig: plasmoid_amount_alt}, continuous reconnection produces around $5-10$ plasmoids constantly during the whole simulation time. A similar scenario also happens in the case $\tt M6a$.
As shown in the 3rd and 4th panels in Fig.~\ref{fig: plasmoid_amount_alt}, in early simulation time ($t \le 4000\, \rm M$), magnetic flux  gradually increases and  then decreases. Such transitions are seen at least two times clearly. 
At the transition from increase to decrease, the polarity of the magnetic field near the horizon starts to change, which leads to strong magnetic reconnection and produces a larger amount of plasmoids. 
Owning to the intermediate size and stronger magnetic field of the loops in the case of $\tt M50a$, at $t \sim 3000\,{\rm M}$ about 15 plasmoids are formed, which is the largest number of plasmoids among the alternating cases. 
In the later transition phase to the MAD state, there is less magnetic reconnection, thus only a few plasmoids are generated.
%
This is because, even in later simulation time, the magnetic loops are still maintained inside the torus by MRI and there is some reconnection between the small-scale magnetic loops. 
The case $\tt M80a$ keeps more large-scale loops and has a longer period of a large number of plasmoid formations. 
From $t=7000$ to $12000\,{\rm M}$, the magnetic field in the torus mixes up and causes frequent reconnection, which leads to plenty of plasmoid formation. After $t=12000\,{\rm M}$, MAD phase starts. During this period, reconnection becomes less frequent, as seen in the case of $\tt M50a$. A larger-scale magnetic field avoids continuous plasmoid formation via frequent reconnection. However, as the polarity in the funnel region still changes occasionally,  there is some plasmoid formation during a such transition period, which corresponds to the spikes in the time evolution curve of the plasmoid amount. 
%

\begin{table*}
\centering
\begin{tabular*}{0.85\textwidth}{@{\extracolsep{\fill}}llllll}
\hline
\multicolumn{1}{l}{Case}             &Average Range & Turbulent, $R_{\rm low}$ & Turbulent, $R_{\rm high}$ & Reconnection, $R_{\rm low}$ & \multicolumn{1}{l}{Reconnection, $R_{\rm high}$} \\ 
\hline
$\tt M0$          & 6000-10000\,M & 1.59  & 6.66  & 2.62  & 6.69  \\
$\tt M6a$         & 5000-9000\,M & 2.11  & 3.63  & 1.94  & 3.87   \\
$\tt M20a$        & 8000-12000\,M & 1.03  & 4.24  & 1.45  & 3.76  \\
$\tt M80a$ (SANE) & 6000-10000\,M & 1.05  & 2.61  & 2.54  & 4.24  \\
$\tt M80a$ (MAD)  & 20000-24000\,M & 2.00  & 5.00  & 2.42  & 3.46  \\
$\tt M6b$         & 5000-9000\,M & 2.54  & 4.78  & 1.11  & 5.93  \\
$\tt M20b$        & 6000-10000\,M & 0.87  & 4.70  & 2.16  & 5.28  \\
$\tt M80b$ (SANE) & 6000-10000\,M & 0.96  & 2.45  & 2.51  & 4.09  \\
$\tt M80b$ (MAD)  & 18000-22000\,M & 1.25  & 5.35  & 2.36  & 4.07  \\
\hline
\end{tabular*}
\caption{Best fit values for $R_{\rm low}$ and $R_{\rm high}$ of turbulent and reconnection heating models respectively.}
\label{Table: R_beta_fitting}
\end{table*}
\begin{figure*}
\centering
	\includegraphics[height=.45\textwidth]{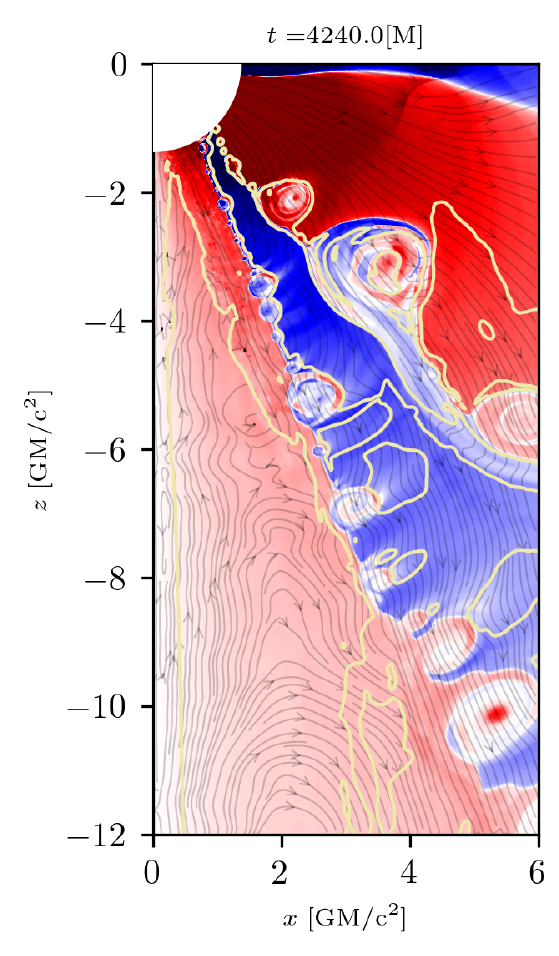}
	\includegraphics[height=.45\textwidth]{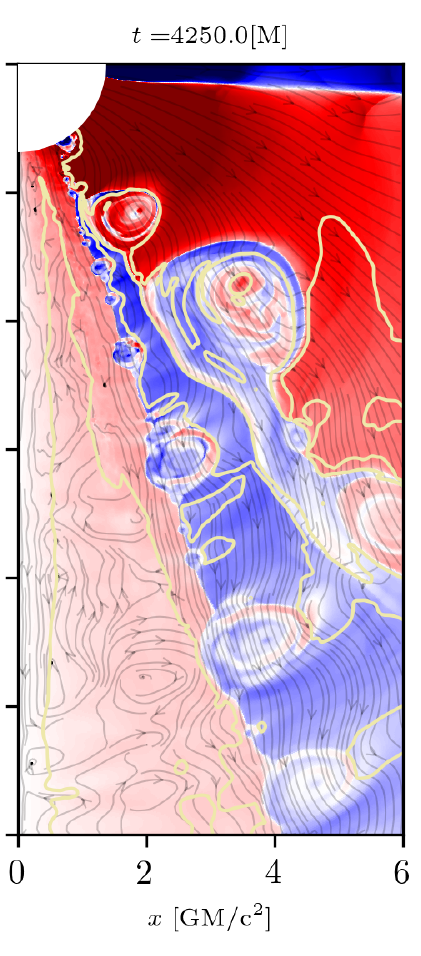}
	\includegraphics[height=.45\textwidth]{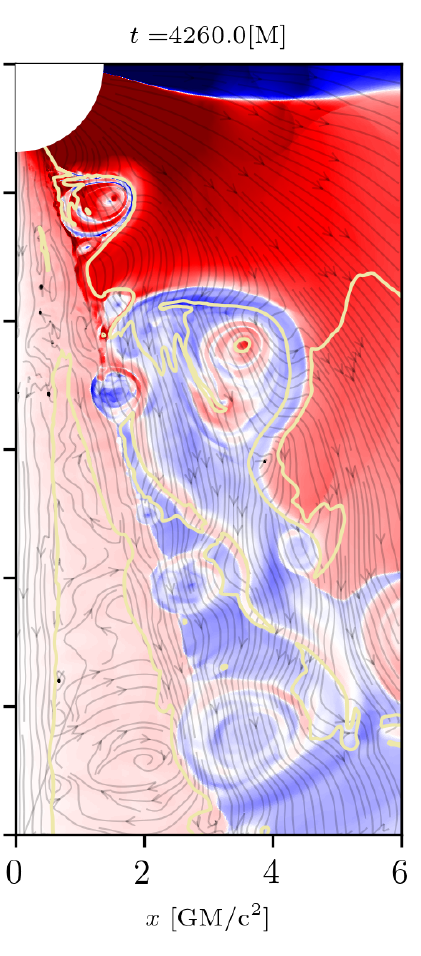}
	\includegraphics[height=.45\textwidth]{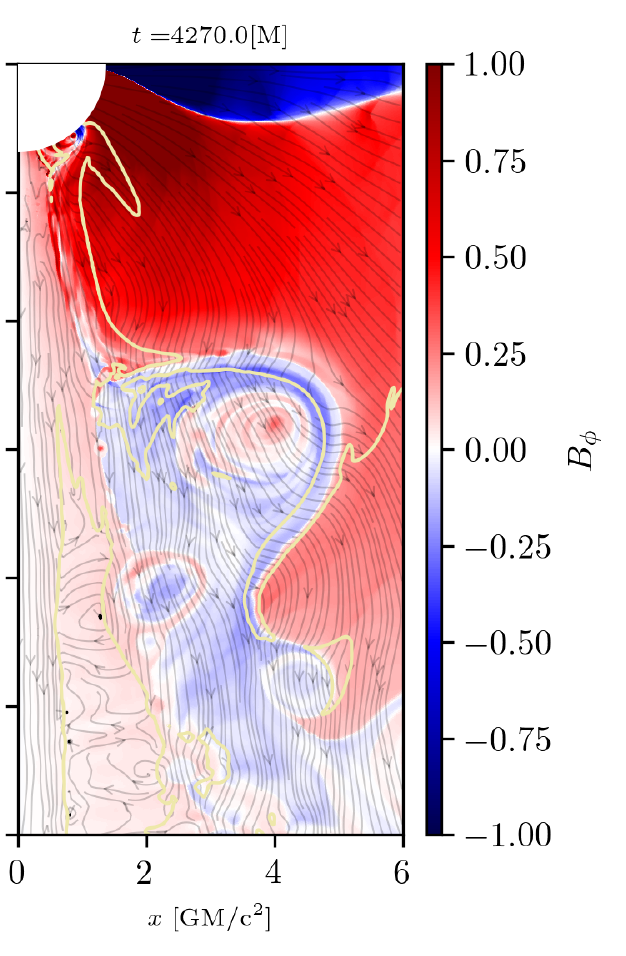}
    \caption{Time evolution of the distribution of the toroidal magnetic field near the horizon region in the high-resolution run of {\tt M80a} case. The streamlines represent the poloidal velocity. Yellow contours in each panel show the transition of gravitationally bounded/unbounded region via the geometric Bernoulli parameter $u_{\rm t}=-1$. We see the development of tearing instability in the current sheet and the formation of plasmoids.
    }
    \label{fig: tearingI}
\end{figure*}

Figure~\ref{fig: plasmoid_amount_sam} shows the plasmoid amount evolution for the same polarity cases with different loop wavelengths which have very different behavior from the alternating ones.
For the case $\tt M6b$, in early simulation time, magnetic dissipation occurs between the loops inside the torus by the MRI due to the short wavelength of the loops as compared with {\tt M6a}. 
However, the reconnection of magnetic loops inside the torus can not form many plasmoids.
When the magnetic field are  strong enough, by the field amplification, there is plasmoid formation around $t=8000\,\rm M$.
After dissipation finishes, the plasmoid formation slows down again.
Case $\tt M20b$ shows similar plasmoid amount evolution as case $\tt M6b$. However, the stronger magnetic field is harder to be fully dissipated. Therefore, the constant reconnection inside and between the loops develops a lot of plasmoids constantly. As discussed in the previous section, dissipation in case $\tt M80b$ is weaker. However, there are still a few plasmoids formed before the accretion flow transits into the MAD state. During the MAD state, plasmoid formation is almost quenched.


%
%
The amounts of plasmoids in the non-rotating black hole cases are significantly lower than that in the rotating black hole cases because most of the plasmoids are gravitationally bounded which makes them hard to be identified with our detection method. As discussed in the previous section, black hole spin has little influence on the accretion process. However, non-rotating black holes have no ergosphere, which leads to less powerful jets \citep{1977MNRAS.179..433B}. Inside the torus, there is less reconnection which results in fewer plasmoids from the tearing instability.

In the case of large magnetic loops with alternating polarity, $\tt M80c$, due to the lower flow velocity difference, the excitement of KH instability is also weak. Relatively strong current sheet and magnetic reconnections are formed at the boundary between the funnel and sheath region which are the main factors of the formation of plasmoids. However, in the case of the same polarity loops, $\tt M80d$ is just the opposite. The reconnection at the jet sheath region becomes weaker. However, due to the magnetic field accumulation on the horizon, it has a stronger outflow in the funnel region. Thus, the developed KH instability becomes stronger. The plasmoid formation rate in these two cases is similar though the reason is different. The situation is similar in the smaller loop cases. 
Due to the lack of the amplification of the magnetic field by the frame-dragging effect, the KH instabilities are significantly weaker, and lower magnetization in the funnel region also limits the strength of the current sheet which leads to the reduction of the number of plasmoids compared with those in the rotating black hole cases.

Previous studies \citep{Ripperda2021,2009PhPl...16k2102B} pointed out high Lundquist number is required to induce tearing instability for plasmoid formation. 
\begin{equation}
    S = v_{\rm A} L/\eta_{\rm num} \gtrsim 10^4,
\end{equation}
where $v_{\rm A}\sim c$ is Alfv\'en speed, $L$ is the length of a current sheet, and $\eta_{\rm num}$ is the resistivity. In our ideal GRMHD simulation, resistivity is induced numerically, and $\eta_{\rm num}\propto \Delta x^p$ with $p\approx 2$ in our second-order calculations in the {\tt BHAC} code \citep{Ripperda2021, Porth2017}. The lengths of the current sheets in our simulations are usually $\gtrsim 5\,\rm r_{\rm g}$ (see Fig.~\ref{fig: tearingI}). Our simulations show that most of the tearing instability happens close to the horizon, in which spatial resolution is approximately $0.025\,\rm r_{\rm g}$ for standard runs. Therefore, as long as the current sheet is longer than $\sim 5\,\rm r_{\rm g}$, tearing instability is likely to occur. For our high-resolution run, $\Delta x$ is 4 times lower, which provides a magnitude higher Lundquist number and stronger tearing instability. Although different numerical resistivity has some minor effect on the plasmoid formation rate, the basic conclusion is not changed. Tearing instability generates a large number of plasmoids at a few gravitational radii away from the black hole in the sheath region.

Fig.~\ref{fig: tearingI} shows the development of the tearing instability and plasmoid generation from our high-resolution run of {\tt M80a} case. Multiple small plasmoids are generated at the current sheet which is close to the black hole ($\lesssim 5\,\rm r_{\rm g}$) (see the first panel of Fig.~\ref{fig: tearingI}). The merger and growth of the plasmoids are presented from the time evolution in Fig.~\ref{fig: tearingI}. Most of the plasmoids are unbounded and moving outward. In the sheath region, two current sheets are seen. The small plasmoids generated from the left one grow and merge over time. A merger in the process of two large plasmoids is shown in the fourth panel of Fig.~\ref{fig: tearingI}.

\begin{figure}
   \centering
        \includegraphics[width=\linewidth]{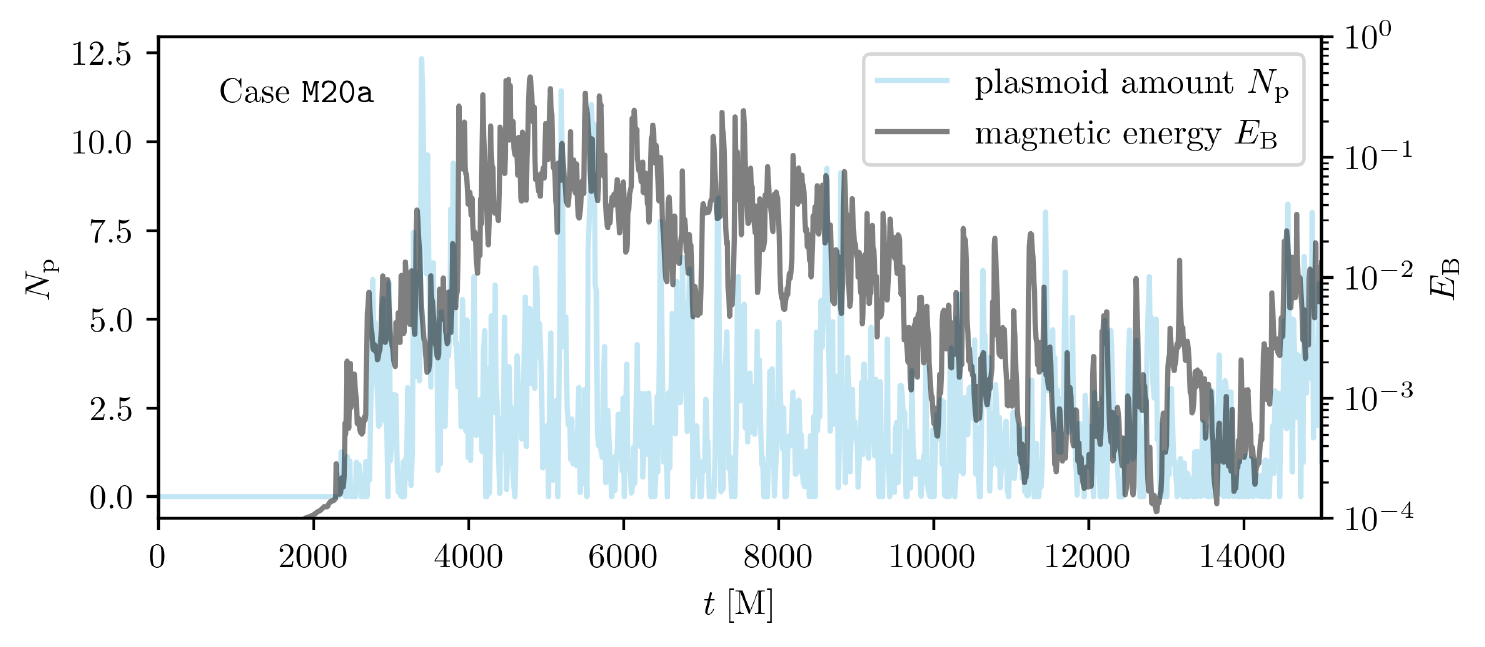}
   \caption{  Time evolution of density-weighted averaged magnetic energy (black) as well as the number of plasmoids (cyan) from case {\tt M20a}.
   }
   \label{fig: B_energy}
\end{figure}

Overlaying the plasmoid amount from Fig.~\ref{fig: plasmoid_amount_alt}, Fig.~\ref{fig: B_energy} presents the density-weighted averaged magnetic energy $E_{\rm B}$ of {\tt M20a} case, which is defined as follows
\begin{equation}
    E_{\rm B} = \frac{B^2\rho}{\bar \rho},
\end{equation}
where $\bar{\rho}$ is averaged density. Plasmoids are mostly generated close to the black hole. We calculate the averaged magnetic energy within the region of $x<30\,\rm r_{\rm g}$ and $-30\,{\rm r_{\rm g}}<y<30\,{\rm r_{\rm g}}$. 
In the time evolution of the simulations, MRI increases the global magnetic energy from approximately $2000-6000\,\rm M$. At the time around $t=3500\,\rm M$, a magnetic loop with different polarity accretes into the black hole causing violent reconnection and plasmoid formation. We see the sudden decrease of the magnetic energy and the peak in the number of plasmoids at this moment in Fig.~\ref{fig: B_energy}). In the following time, a similar anti-correlation between magnetic energy and a number of plasmoids demonstrates the relation between reconnection and plasmoid formation.

As suggested in \cite{2022ApJ...929...62B}, both KH and tearing instabilities are responsible for the formation of plasmoids. From the simulations, we see some plasmoids are formed by KH instability alone. 
As shown in the left panel of Fig.~\ref{fig: KHI and reconnection}, the streamlines of the velocity field show clear vorticity around the plasmoid which is a sign of KH instability. From the time evolution of the simulations, some of the plasmoids developed by KH instability have very fast rotation (vorticity). The poloidal velocity distribution also shows strong velocity shear at the plasmoid location seen in the left panel of Fig.~\ref{fig: KHI and reconnection}. On the other hand, the plasmoids formed in current sheets by tearing instability and magnetic reconnection do not show explicit rotating behavior or vortex field (see the right panel of Fig.~\ref{fig: KHI and reconnection}). 
We compare the electron temperatures of plasmoids produced between KH instability and tearing instability. The electron temperature of plasmoids made by KH instability is lower than those made by tearing instability. Generally, the lower electron temperature is likely to make them harder to be recognized from the observation. 

%
In rotating black hole cases, the plasmoids are usually created at the funnel wall (the boundary between the jet funnel and sheath) and advected along the jets. They become bigger and dissipate their energy gradually. However, due to the weak outflow, plasmoids in the non-rotating black hole cases are not pushed out. They are wandering in the sheath region. That different trajectory of plasmoids leads to the dynamical difference in different black hole spin.

\section{Electron temperature properties}

\begin{figure*}
	\includegraphics[height=0.4\textwidth]{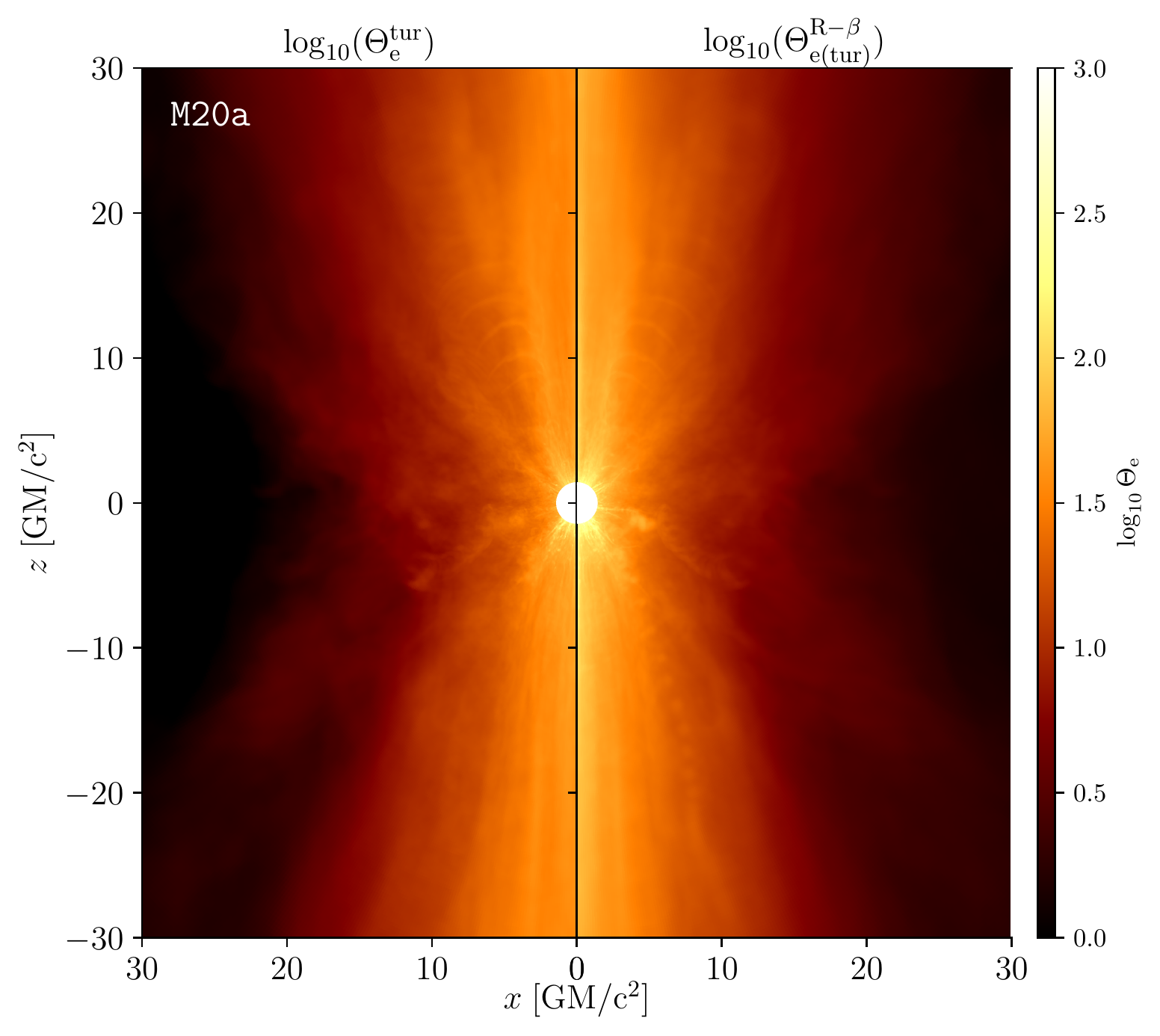}
	\includegraphics[height=0.4\textwidth]{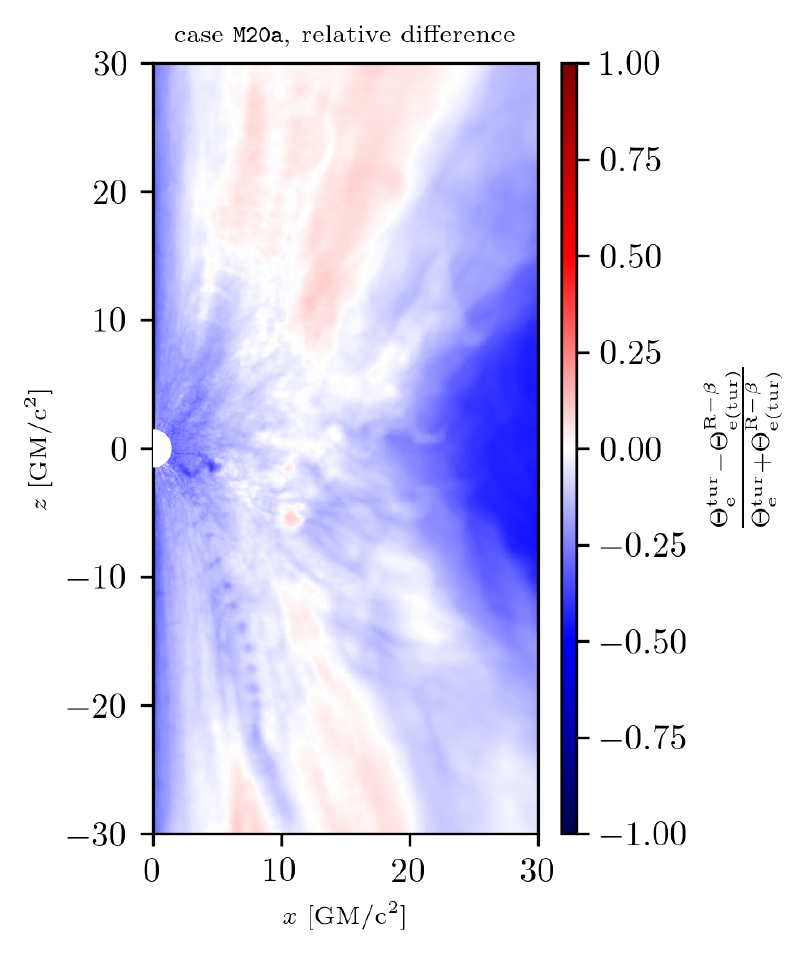}
    \caption{The distribution of ({\it left}) time-averaged normalized electron temperature calculated from turbulent heating prescription, ({\it middle}) that from parameterized $R-\beta$ prescription whose parameters are fitted from the two-temperature model ($R_{\rm low}=1.03$ and $R_{\rm high}=4.24$), and ({\it right}) the relative difference of two-temperature simulation and $R-\beta$ model for the case of rotating black hole with alternating polarity $\tt M20a$, respectively. The average time range is from $t=8000$ to $12000\,M$. }
    \label{fig: averaged_electron_temperature}
\end{figure*}

\begin{figure*}
    \centering
         \includegraphics[width=\linewidth]{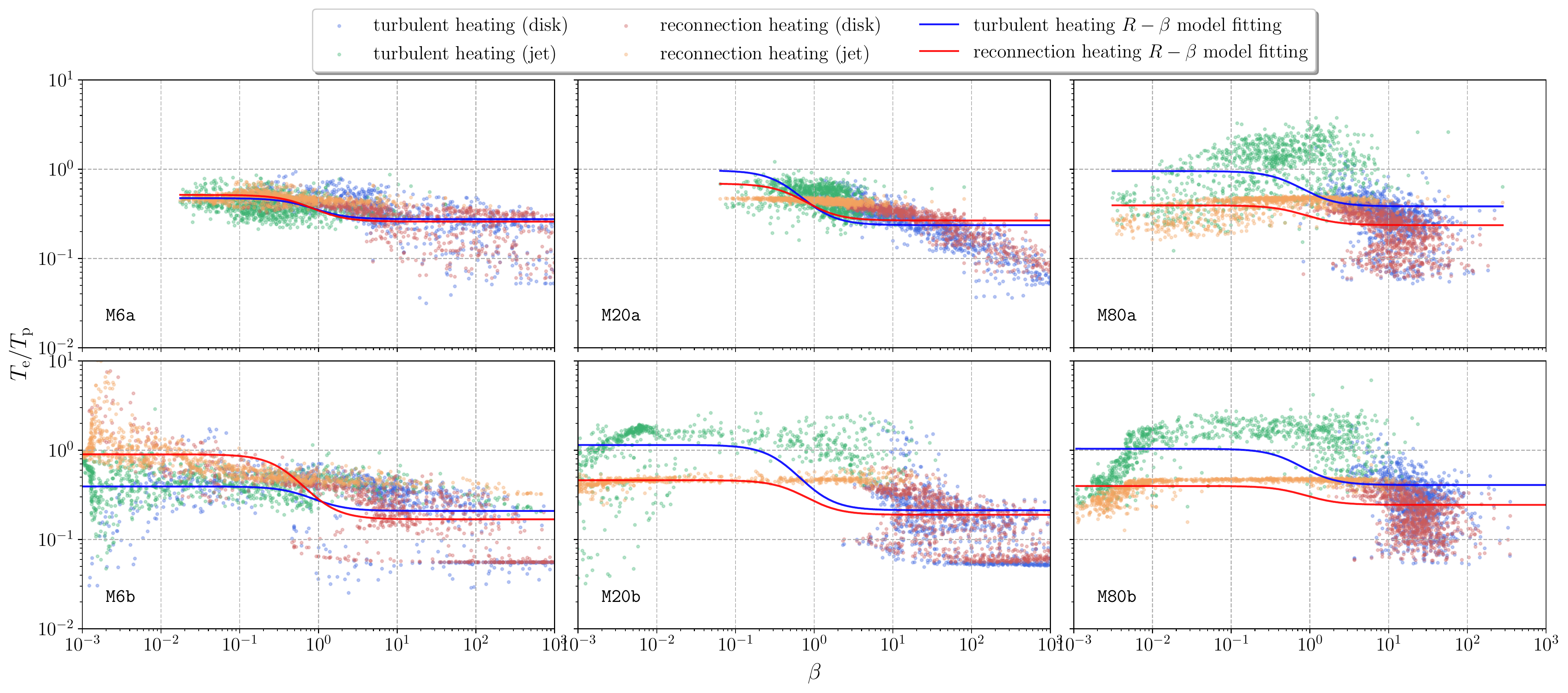}
    \caption{$T_{\rm e}/T_{\rm p}-\beta$ diagram for the multiple loops configuration cases ({\it upper}: alternating polarity and {\it lower:} same polarity) with different loop wavelengths for rotating black holes. Red (blue) and green (orange) points indicate the disk and jet components in turbulent (magnetic reconnection) heating prescription, respectively. Scattered points are taken from the time-averaged data of $t=4000\,M$ time range within $r \le 100 \, M$. Each selected averaged time range is summarized in Table~\ref{Table: R_beta_fitting}. Note that for the cases $\tt M80a$ and $\tt M80b$ there is both SANE and MAD regime during time evolution. We select the time on the SANE regime. Blue and red solid lines are best-fit curves using R-$\beta$ parametrized prescription. }
    \label{fig: TeTp-beta}
\end{figure*}
\begin{figure*}
    \centering
         \includegraphics[width=\linewidth]{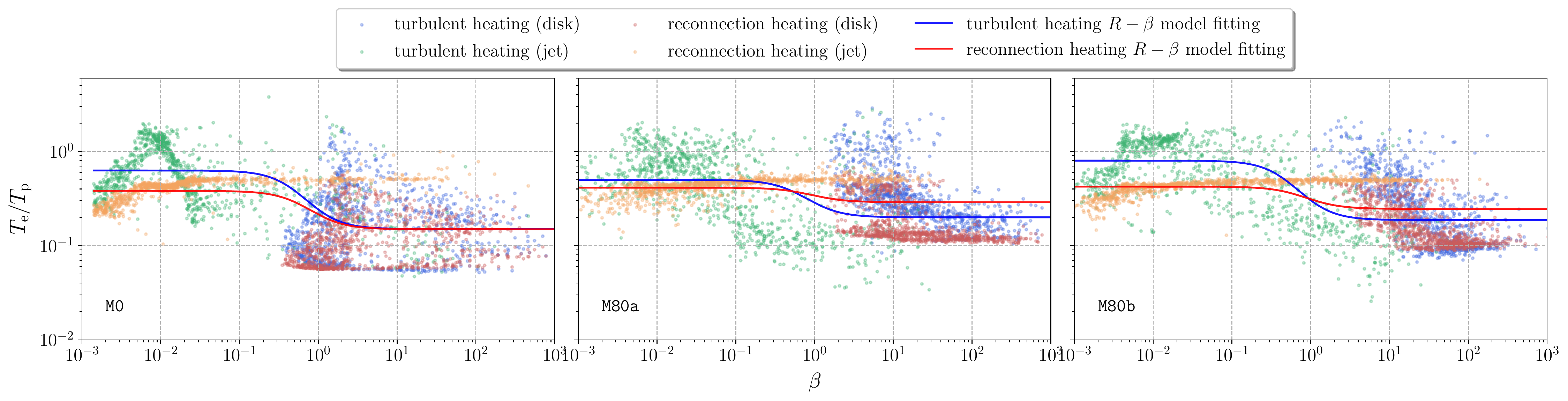}
    \caption{Same as Fig.~\ref{fig: TeTp-beta} but in the simulation time during the MAD state. The different panels show  {\it left:} single loop case $\tt M0$, {\it middle:} alternating polarity multi-loop case $\tt M80a$ and {\it right:} same polarity multi-loop case {\tt M80b}, respectively.}
    \label{fig: TeTp-beta_MAD}
\end{figure*}

Electron temperature is an important quantity for the connection to radiation. From our two-temperature GRMHD simulations, we can obtain electron temperature directly from simulations.
Figure~\ref{fig: averaged_electron_temperature} shows the time-averaged normalized electron temperature distribution by the turbulent heating prescription in two-temperature simulations and the $R-\beta$ parameterized prescription using the same GRMHD data along with their relative difference. 
The expression of the parameterized $R-\beta$ prescription is given as follows
\begin{equation}
    \frac{T_{\rm p}}{T_{\rm e}} = \frac{1}{1+\beta^2}R_{\rm low}+\frac{\beta^2}{1+\beta^2}R_{\rm high},
    \label{Eq: R-beta}
\end{equation}
where $R_{\rm low}$ and $R_{\rm high}$ are the free parameters of the model. For the $R_{\rm low}$ and $R_{\rm high}$ values in the $R-\beta$ model, we use the best-fit values from the results of the distribution of electron-to-ion temperature ratio obtained from our two-temperature simulations (see Fig.~\ref{fig: TeTp-beta}). 
The overall normalized electron temperature distributions are similar in both cases, it is higher in the funnel region, implying a bright jet compared with the accretion disk region. However, the turbulent heating one shows a more small patchy high electron temperature structure, while the result of the $R-\beta$ model has a more extended distribution. As a consequence, in order to better extract the local current sheet or other small high electron temperature structure, the two-temperature calculation may produce a more accurate result.
From the pixel-by-pixel comparison, the normalized electron temperature by the turbulent heating prescription has a higher value than in the $R-\beta$ model in the funnel and shearing boundary region. On the other hand, in the disk region, the normalized electron temperature by the turbulent heating prescription is lower than in the $R-\beta$ model. 
We should note that the time-averaged figures of turbulent heating and reconnection heating are very similar, therefore, we only put the result of the turbulent heating in Fig.~\ref{fig: averaged_electron_temperature}.

\begin{figure*}
	\includegraphics[height=0.4\textwidth]{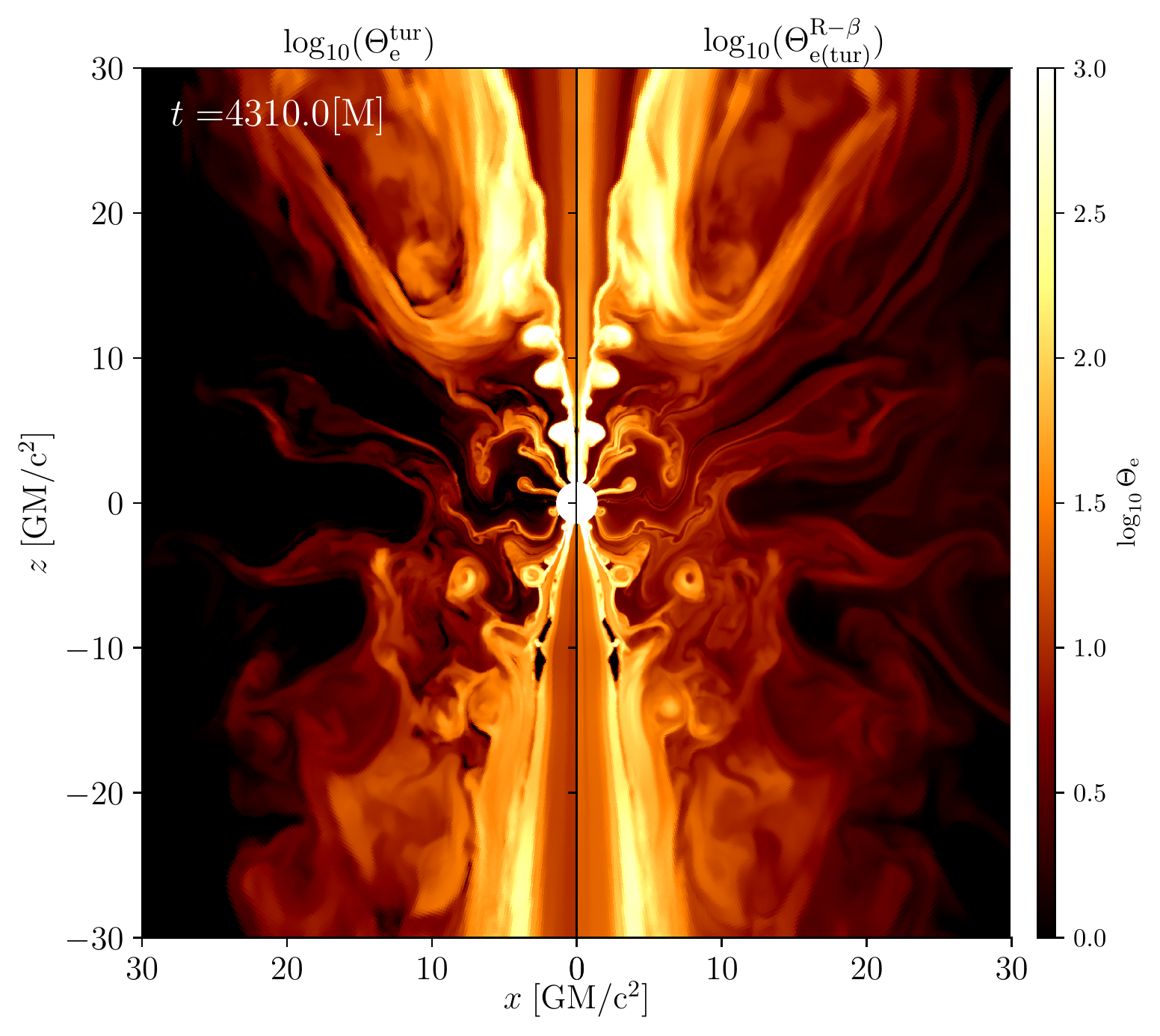}
	\includegraphics[height=0.4\textwidth]{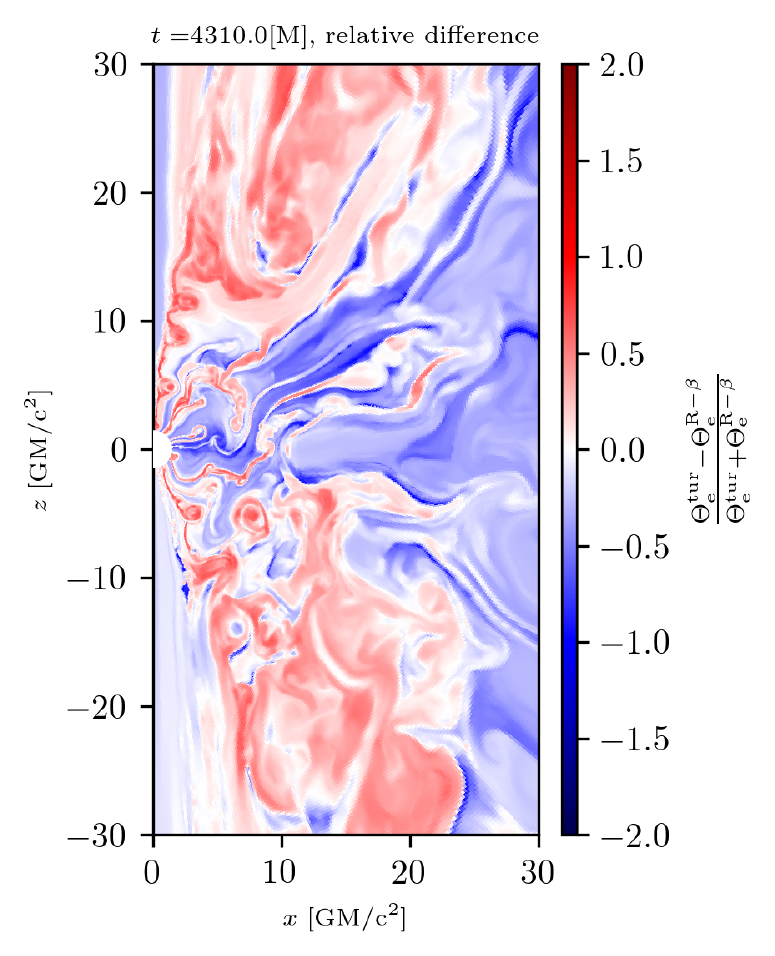}
    \caption{Same with Fig.~\ref{fig: averaged_electron_temperature} but using snapshot data from the case of rotating black hole with alternating polarity $\tt M20a$ at $t=4310\,M$.}
    \label{fig: plasmoid electron temperature}
\end{figure*}


In Fig.~\ref{fig: TeTp-beta}, the relationships between electron-to-ion temperature ratio $T_{\rm e}/T_{\rm i}$ and plasma $\beta$ for the cases of multiple loops configuration during the SANE state are presented. The values correspond to the averaged data of $t=4000\,M$ time range. 
The averaged time range for each case is summarized in Table~\ref{Table: R_beta_fitting}. 
The scatter plots for both turbulent and magnetic reconnection heating prescriptions are well-fitted with $R-\beta$ parameterized prescription.
As we see in Fig.~\ref{fig: TeTp-beta}, the relationship between $T_{\rm e}/T_{\rm i}$ and plasma beta are varied with the different magnetic field configurations and black hole spins.
Generally, both electron and ion temperatures become higher in low plasma $\beta$ regions. In some cases, in the very low plasma beta region, the electron temperature exceeds the ion temperature. 
In single fluid GRMHD simulations, in order to obtain electron temperature from $R-\beta$ parameterized prescription, we usually assume $T_{\rm e} \approx T_{\rm i}$. Therefore, $T_e/T_i =1$ is the maximum. Such high electron temperature in the low $\beta$ region corresponds to the funnel region with high magnetization. These regions are typically affected by numerical floors. Therefore, the electron temperature value is overestimated \cite[e.g.,][]{Mizuno2021}.


In previous works \citep[e.g.,][]{Moscibrodzka2016,EHTpaperV,Mizuno2021}
$R_{\rm low}$ is fixed to be 1 and $R_{\rm high}$ is set to be a free parameter.
In this work, we set both $R_{\rm low}$ and $R_{\rm high}$ as free parameters. 
We perform a least square fitting in each case and best-fit parameters are shown in Table~\ref{Table: R_beta_fitting}. 
From the best fit vales of $R_{\rm low}$ and $R_{\rm high}$, both ranges are roughly $1 < R_{\rm low} < 3$ and $5< R_{\rm high} < 10$. These best-fit values are similar to the results in single loop cases \citep{Mizuno2021}. The results indicate that the magnetic field configuration does not much affect the electron heating properties.

In general, the turbulent heating model has a lower $R_{\rm low}$ value than that in magnetic reconnection heating prescription. This discrepancy is seen more clearly during the MAD state. Figure~\ref{fig: TeTp-beta_MAD} shows the scatter plot of electron-to-ion temperature ratio as a function of plasma beta in three different cases, single loop case of $\tt M0$, multiple-loops case with alternating polarity $\tt M80a$, and multiple-loops case with same polarity $\tt M80b$. They are time-averaged during MAD state.  
In low $\beta$ regions where it is usually located in the funnel region, the electron temperature of turbulent heating prescription is higher than reconnection heating, while in high $\beta$ regions, the reconnection heating is higher than turbulent heating. It is because of the maximum efficiency of each heating prescription. In turbulent heating, maximum efficiency $f_{\rm e} \sim 1$ but for magnetic reconnection heating the maximum is $0.5$ (see Eq.~\ref{eq2} and Eq.~\ref{Eq: reconnection heating}). Thus, a factor 2 difference is reflected in particular, in low $\beta$ regions.
This difference implies turbulent heating tends to produce a brighter jet in radiation images and reconnection makes the accretion disk region brighter. In reality, both heating mechanisms co-exist. By carefully comparing  radiation images by general relativistic radiation transfer (GRRT) calculation and observation, we may determine the contribution of each heating mechanism. However, our calculation does not include the effect of radiative cooling. Therefore, the calculated temperature here would be somehow overestimated \citep[][]{Ryan2018, Chael2018, Yoon2020, Dihingia2023}. 
 


\subsection{Comparison of two-temperature and $R-\beta$ model in plasmoids}

As discussed in the previous section, compared with the $R-\beta$ model, the electron temperature distribution of the two-temperature models has higher electron temperature at funnel regions. It reflects the superiority of two-temperature models in dealing with higher electron temperature regions and the underlying physics. As seen in Fig.~\ref{fig: plasmoid electron temperature}, the turbulent heating model clearly shows higher contrast than the $R-\beta$ one. For the torus region, the two-temperature model gives lower electron temperature, which makes higher variability. For the current sheet and plasmoids, the two-temperature model has a significantly higher electron temperature. 

\begin{figure}
    \centering
	\includegraphics[height=1\columnwidth]{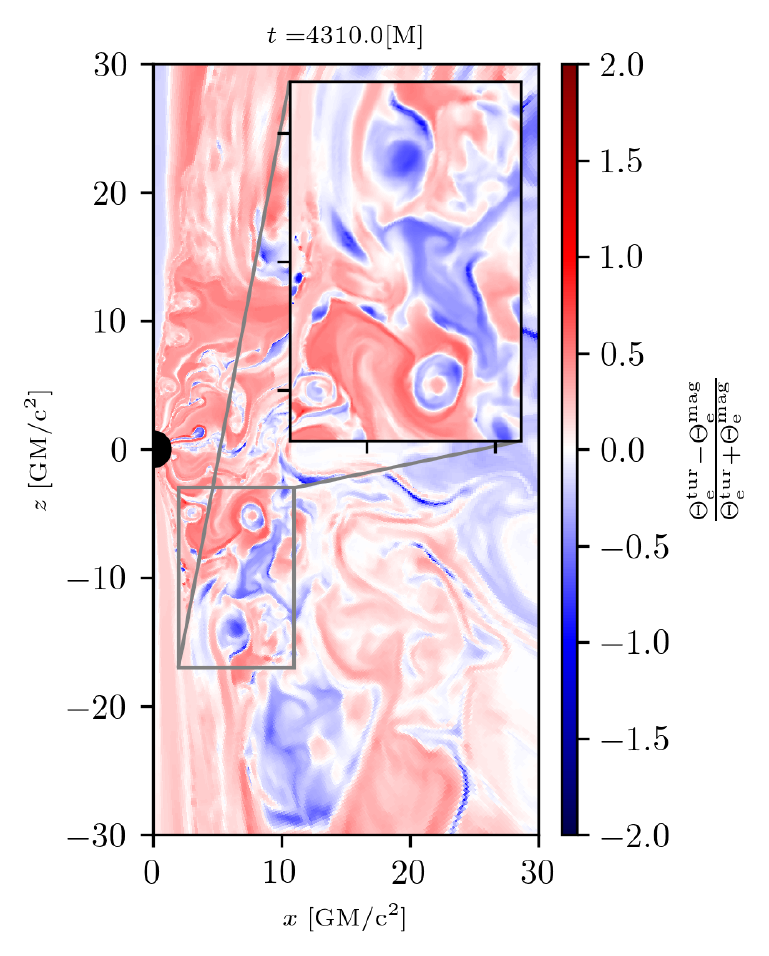}
    \caption{Relative difference of the turbulent and reconnection heating models at the plasmoid related region taken from the case of rotating black hole with alternating polarity $\tt M20a$ at $t=4310\,M$.}
    \label{fig: difference_plasmoid}
\end{figure}

The two heating models used in this work, reconnection heating and turbulent heating, also have some differences in plasmoids. Note that inside the plasmoids, the distribution of the magnetic field shows clear three layers, i.e., a high-density core with a strong magnetic field, a small current sheet, and an outer layer with a weaker magnetic field. This structure is more obvious when the plasmoid is related to KH instability (see the left panel of Fig.~\ref{fig: KHI and reconnection}), while for the plasmoids from tearing instability, such a high-density core is not visible. Generally, compared with reconnection heating, turbulent heating gives higher electron temperature in the strong magnetic regions (see Fig.~\ref{fig: TeTp-beta} and \ref{fig: TeTp-beta_MAD}). Turbulent heating produces higher electron temperature in the funnel and most high-temperature regions, including plasmoids from tearing instability. However, from Fig.~\ref{fig: difference_plasmoid} we observe that the reconnection heating model produces higher electron temperature inside the KH instability-related plasmoids. Combining Fig.~\ref{fig: difference_plasmoid} and the $R-\beta$ model fitting results in Figs.~\ref{fig: TeTp-beta} and \ref{fig: TeTp-beta_MAD}, we also conclude that in general, the turbulent heating model gives higher electron temperature than the reconnection heating one in the regions with a higher magnetic field (e.g., jet, plasmoid), while for the low magnetic field regions such as the accretion torus, reconnection heating gives higher electron temperature. 

\begin{figure}
    \centering
    \includegraphics[width=\columnwidth]{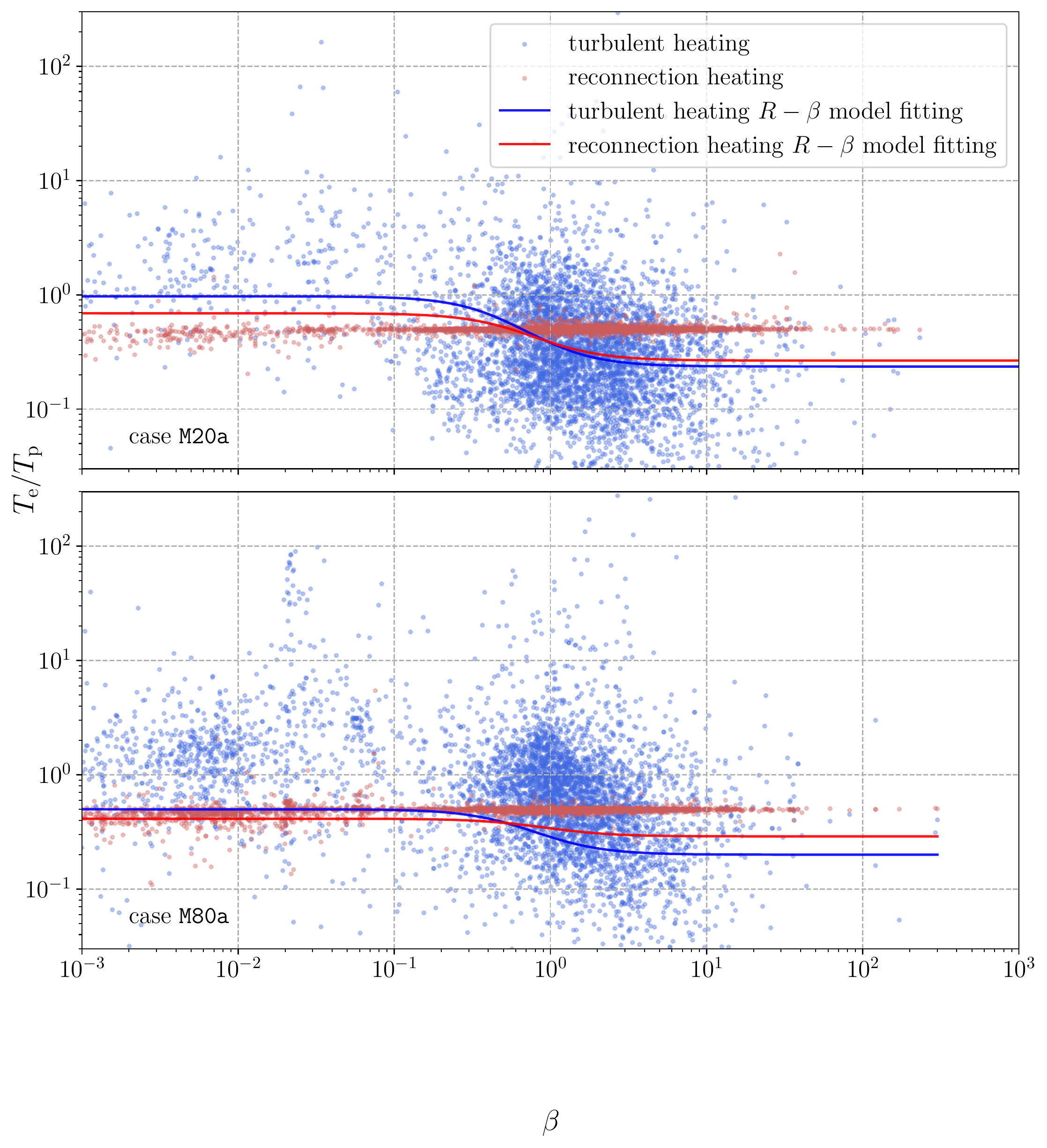}
    \caption{Same as Fig.~\ref{fig: TeTp-beta} but data points are collected from the cells where plasmoids locate all snapshots. Fitting parameters are from Table~\ref{Table: R_beta_fitting}.}
    \label{fig: plasmoid_scatter}
\end{figure}
Fig.~\ref{fig: plasmoid_scatter} shows the $T_{\rm e}/T_{\rm p}-\beta$ diagram of the plasmoid regions of two representative cases, $\tt M80a$ and $\tt M20a$. The scatter points of the reconnection heating are heated up to the maximum allowed by Eq.~\ref{Eq: reconnection heating}, while the turbulent heating case shows more variability. The variability of the latter one suggests some plasmoids do not get sufficient heating. Thus, reconnection heating provides more heat on plasmoids. The parameters of the $R-\beta$ model used in this figure come from the fitting result in Table~\ref{Table: R_beta_fitting}. For case $\tt M80a$ we use the parameters calculated during the MAD phase. $R_\beta$ model fits the distribution of reconnection heating model well when $\beta<1$. However, it fails for the turbulent heating model and the $\beta>1$ part of the reconnection heating model.  
\begin{figure}
    \centering
         \includegraphics[width=\linewidth]{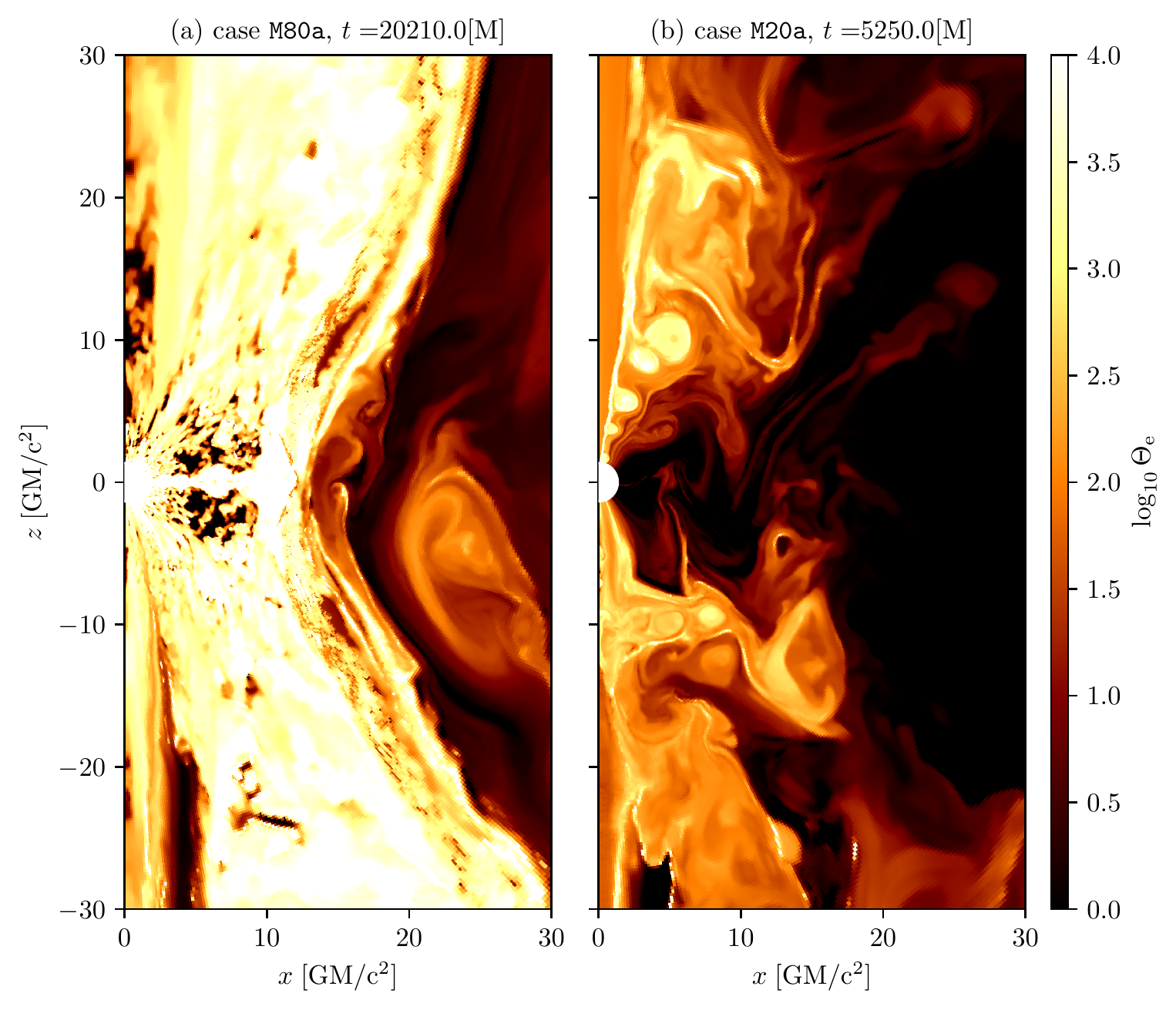}
    \caption{Distribution of normalized electron temperature for {\it left} the representative of accretion flows in MAD regime ($t=20210\,\rm M$ of the case $\tt M80a$) and  {\it right} the representative of accretion flows in SANE regime ($t=5250\,\rm M$ of the case $\tt M20a$).}
    \label{fig: jet_plasmoid}
\end{figure}

As shown in Figure~\ref{fig: jet_plasmoid}, accretion flows during the MAD regime produces strong jets with high electron temperature $\Theta_{\rm e}$. They would over-shine the plasmoids. Although the plasmoids have relatively higher electron temperatures than other cases, because of the existence of the strong jet, the effect of plasmoids would be hidden in the emission. However, in the case $\tt M20a$ there is no clear strong jets in the funnel region. The plasmoids have the highest electron temperature. The radiation should come from the hot plasmoids and it would dominate the variability of the light curves. 
In our future work, we will investigate the light curves from radiation images obtained from the GRMHD simulations with the multiple loops configuration and discuss the connection to the plasmoid formation.

\section{Summary and Discussion}

We have performed a series of 2D two-temperature GRMHD simulations initiated with different magnetic field configurations of multiple poloidal loops. For small loop cases such as $\lambda_{\rm r}=6$ ($\tt M6a,b,c,d$) and $\lambda_{\rm r}=20$ ($\tt M20a,b,c,d$), the accretion flows never reach MAD regime. 
In the alternating polarity cases, due to the development of MRI inside the torus, the magnetic field of neighboring loops interact with each other, leading to strong dissipation by magnetic reconnection. 
Such reconnection inside the torus breaks the regularity of loops and leads to the dissipation of most of the magnetic field before accreting onto a black hole. Therefore, even in the long-term evolution, the simulations do not show any MAD-like behavior.
For the same polarity cases, the magnetic reconnection happens at the boundaries between the loops at the early simulation time. 
After the initial dissipation period, roughly one polarity loop survives.

Larger magnetic loop cases show similar behavior in early simulation time but in late time, they have shown the MAD-like signature. 
In particular, for the alternating polarity cases, due to the longer loop wavelength, developed turbulence by MRI does not make much dissipation inside the torus. Thus, alternating polarity loops continuously accrete onto the black hole. Once an opposite polarity loop accretes, magnetic reconnection occurs and a large number of plasmoids form between the two loops. Strong reconnection also interrupts the formation of jets, causing a one-sided jet or only weak outflow. We see the development of stripped jet. 
After enough accumulation of magnetic field on the horizon, the MAD regime shows up. 
The magnetic field strength and mass density of the accretion flow in multiple loop cases are much lower than those in the MAD state of the single loop cases. 
During the time of the MAD state, the plasmoid formation rate becomes decrease. Once opposite polarity loops are accreted again, magnetic reconnection occurs then a lot of plasmoids are formed. The MAD regime is back to the SANE state again and restarts the magnetic field accumulation. The long-term simulations have shown several transitions of the accretion state between SANE and MAD. We found that the transition time depends on the initial loop wavelength.
 

We also found that plasmoid formation rate and their properties strongly depend on the size of the magnetic loops in the initial magnetic field configuration. 
Plasmoid formation in our simulations is highly related to two possible mechanisms, magnetic reconnection, and plasma instabilities, e.g. tearing and KH instabilities. For the rotating black hole cases, the frame-dragging effect of the black hole amplifies the magnetic field in the funnel region, which causes strong reconnection at the boundary region between the jet and sheath. For the shorter loop wavelength cases $\tt M6a,b$, and $\tt M20a,b$, due to the smaller initial magnetic loops, reconnection is more frequent happened than those in the cases with larger loops. 
We found that the plasmoid formation rate in the non-rotating black hole cases is lower than those in the rotating black hole cases. There are several reasons. Lacking the amplification of the magnetic field from the frame-dragging effect of the rotating black hole, the reconnection becomes much weaker. There is no formation of a strong jet. It may imply weaker KH instability due to weaker velocity shear.
In our simulations we observed clear evidence for the coexistence of two plasmoid formation mechanisms, KH instability and reconnection \citep{2022ApJ...929...62B}. The plasmoids formed by the KH instability are typically gravitationally bounded and show lower electron temperatures. The large velocity difference of different fluid layers causes a fast plasmoid rotation. Velocity streamlines show the vortex structure around them. The plasmoids formed by the reconnection have a strong extended current sheet. It forms a plasmoid chain by the tearing instability.
 
The electron temperature is an important physical quantity for the connection to the observed radiation. In our simulations, we have used two different electron heating prescriptions (turbulent and magnetic reconnection) which provide the electron temperature directly from the GRMHD simulations. 
The distribution of electron-to-ion temperature ratio in the simulations has been investigated. We found that commonly used R-$\beta$ parameterized prescription is well-matched in both heating prescriptions even in the different magnetic field configurations as seen in \cite{Mizuno2021}. In the comparison between electron heating prescriptions, turbulent heating leads to higher $R_{\rm low}$ and $R_{\rm high}$ best-fit values.      
In general, the electron temperature is high in the funnel region and plasmoids although we may overestimate them due to lack of radiative cooling \citep{Ryan2018,Chael2018,Yoon2020,Dihingia2023}. If we do not have strong jets in the simulations, plasmoids will be a dominant feature in the radiation which will be a hotspot surrounding a black hole and reflect the variability of light curves as flares seen in Sgr\,A*.
 
Inside the plasmoids, we see the difference in electron temperature between electron hearing models and parameterized $R-\beta$ model. In general, parameterized $R-\beta$ prescription gives a lower estimate of electron temperature at the plasmoids. Between the two-electron heating prescriptions, turbulent and reconnection heating also show some differences. Reconnection heating heats the plasmoids up extensively which forms the roughly single value of electron-to-ion temperature ratio with no dependence on the plasma beta. While the turbulent heating model shows more scattered values of electron-to-ion temperature ratio. 
Such results reflect more detailed physics brought by two-temperature simulations which may not follow by simple parameterized prescription. 
 

Recently \cite{Nathanail2021} has indicated that the magnetic reconnection site in the multi-loops configuration simulations has an intrinsically three-dimensional nature. The magnetic-field lines are sheared and twisted inside the funnel region. Similar scenarios are also seen in our 2D simulations (the right panel of Fig.~\ref{fig: jet_plasmoid}). However, as discussed in \cite{Nathanail2021}, we are lacking the information on $\phi$ direction. The three-dimensional effect makes the plasmoid to be flux ropes twisted around the axis of the torus. Thus, in order to understand the plasmoid properties more precisely, We plan to perform the 3D two-temperature GRMHD simulations with multiple poloidal loops in near future.

\section*{Acknowledgements}
This research is supported by the National Natural Science Foundation of China (Grant No. 12273022), the Shanghai orientation program of basic research for international scientists (Grant No. 22JC1410600), the DFG research grant ``Jet physics on
horizon scales and beyond" (Grant No. FR 4069/2-1), and the ERC Advanced
Grant ``JETSET: Launching, propagation and emission of relativistic jets
from binary mergers and across mass scales'' (Grant No. 884631).
The simulations were performed on Pi2.0 and Siyuan Mark-I at Shanghai Jiao Tong University.

\section*{Data availability}
The data underlying this article will be shared on reasonable request to the corresponding author.





\appendix

\section{Comparison of numerical resolutions}
\begin{figure}
    \centering
    \includegraphics[width=0.9\linewidth]{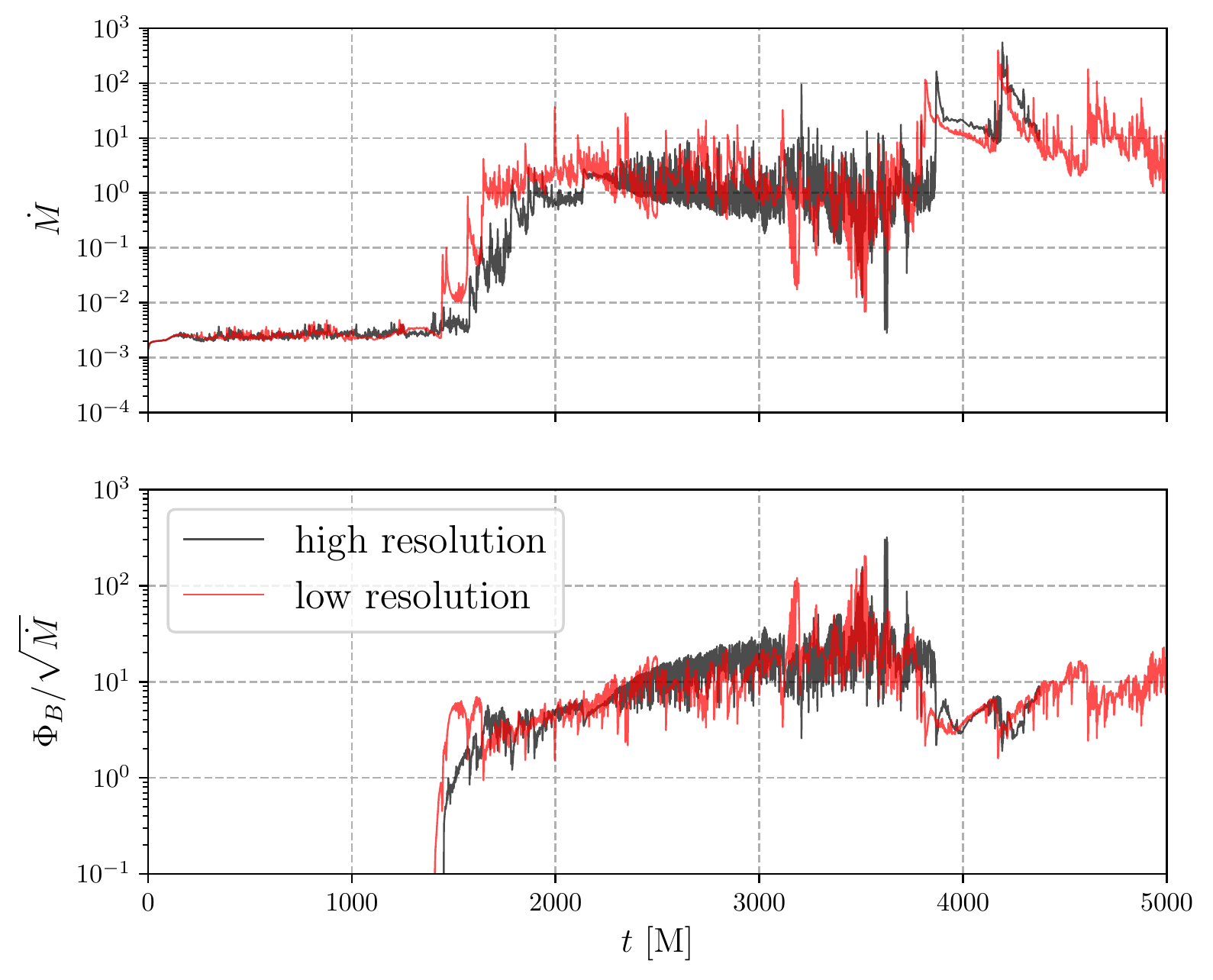}
    \caption{Time evolution of the mass accretion rate ({\it upper}) and the magnetic flux ({\it lower}) onto the horizon for the case of a rotating black hole with long wavelength alternating polarity loops ({\tt M80a}). Black and red lines are high and low resolution, respectively.
    }
    \label{fig: Mdot_high_low_res}
\end{figure}
\begin{figure}
    \centering
    \includegraphics[width=0.9\linewidth]{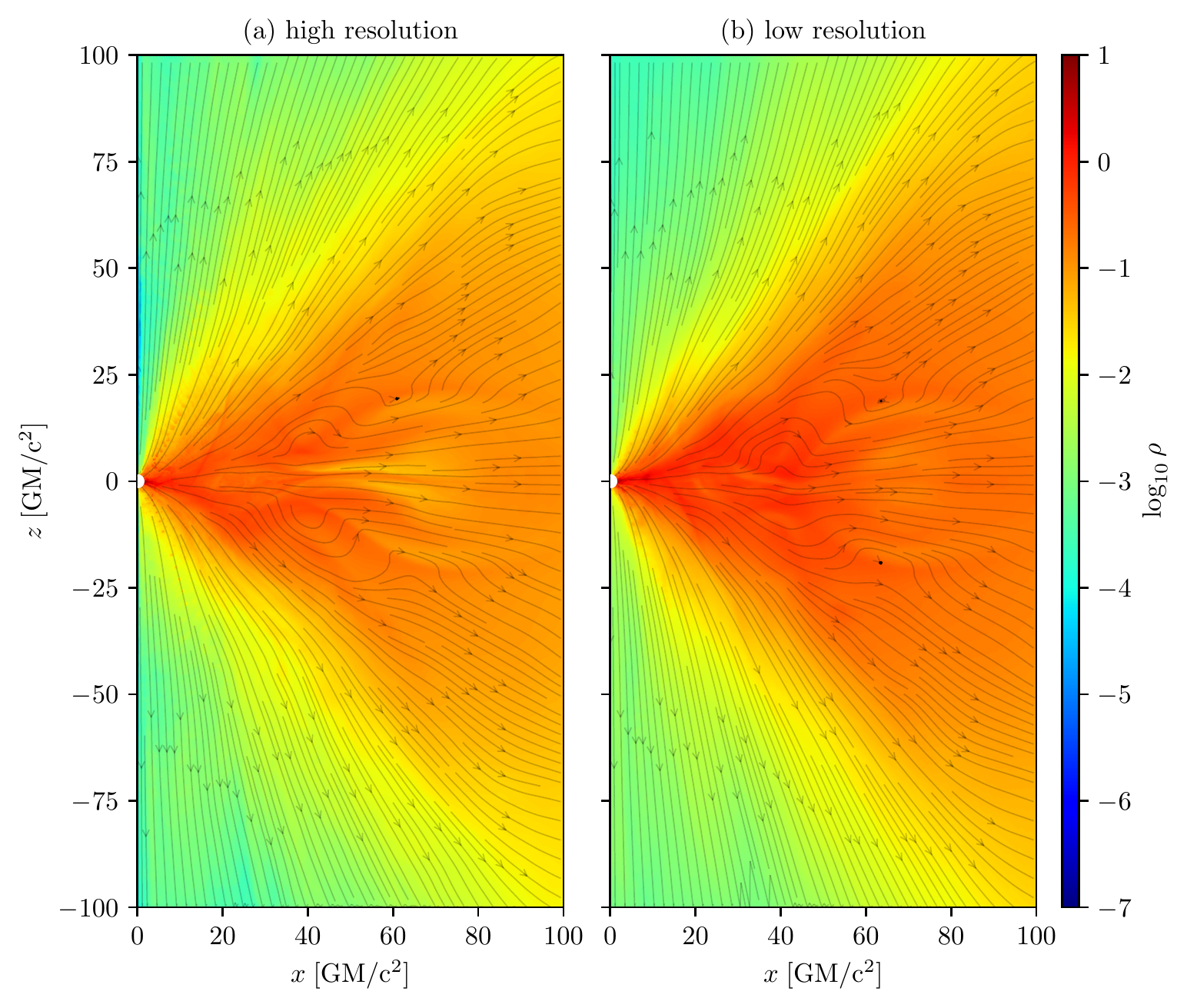}
    \includegraphics[width=0.9\linewidth]{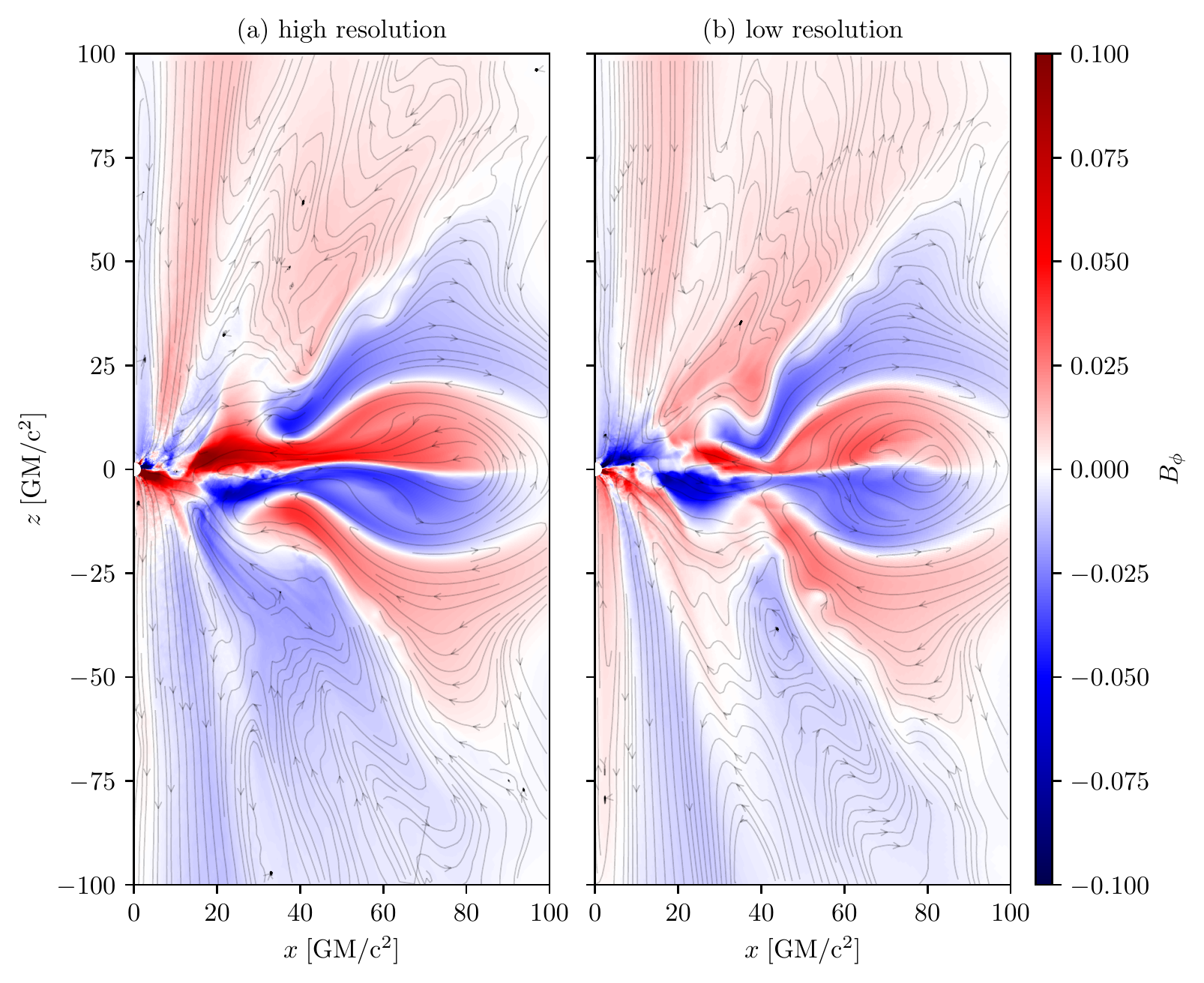}
    \caption{Distribution of time-averaged logarithmic density ({\it upper}) and toroidal magnetic field ({\it lower}) for the case of rotating black hole with long wavelength alternating polarity loops ({\tt M80a}) with high ({\it left}) and low ({\it right}) resolutions. Black streamlines in upper panels represent fluid velocity and those in lower panels show a poloidal magnetic field.}
    \label{fig: 2D_high_low_res}
\end{figure}

In order to investigate the resolution dependency, we perform 2D GRMHD simulations of magnetized accretion flows onto a rotating black hole with multiple alternating polarity loops ($\lambda_r=80$) in different numerical resolutions. We set the same initial conditions for the simulations changing resolution. We choose the grid number, $(N_r, N_\theta) = (1024, 512)$ as a standard resolution and $(N_r, N_\theta) = (4096, 2048)$ as a high resolution. 

Figure~\ref{fig: Mdot_high_low_res} shows the time evolution of mass accretion and magnetic flux rates onto the horizon. The evolutionary trends in both different resolution cases are matched very well. Time-averaged density, velocity and magnetic field distributions are presented in Fig.~\ref{fig: 2D_high_low_res}. Although higher resolution cases show a higher contrast and sharpened sub-structure, these global distributions are similar. Therefore, we conclude that the effect of the numerical resolution is minor in our cases and this lower numerical resolution is enough for the investigation of the properties of accretion flows.

\bsp	
\label{lastpage}
\end{document}